\documentclass[%
a4paper,
twocolumn,
superscriptaddress, 
showkeys,
 amsmath,amssymb,
 aps,
prr,
floatfix
]{revtex4-2}

\usepackage{graphicx}
\usepackage{dcolumn}
\usepackage{verbatim}
\usepackage{bm}

\usepackage{comment}
\usepackage[normalem]{ulem}
\usepackage{float}

\usepackage{comment}

\usepackage{relsize}

\usepackage[hidelinks,unicode=true]{hyperref}
\hypersetup{colorlinks=true,
  linkcolor=blue,
  urlcolor=blue,
  citecolor=blue,
  pdfhighlight=/N
}

\usepackage[caption=false, position=top, font={bf,footnotesize}, justification=raggedright, singlelinecheck=false]{subfig}

\graphicspath{{images_main/}{./images_SI/}}

\usepackage{xcolor}

\newcommand{\av}[1]{\langle{#1}\rangle}

\begin{document}

\title{Signatures of criticality in turning avalanches of schooling fish}

\author{Andreu Puy}
\email{andreu.puy@upc.edu}
\affiliation{Departament de Física, Universitat Politècnica de Catalunya, Campus Nord B4, 08034 Barcelona, Spain}

\author{Elisabet Gimeno}
\affiliation{Departament de Física, Universitat Politècnica de Catalunya, Campus Nord B4, 08034 Barcelona, Spain}
\affiliation{Departament de Física de la Matèria Condensada, Universitat de Barcelona, Martí i Franquès 1, 08028 Barcelona, Spain}

\author{David March-Pons}
\affiliation{Departament de Física, Universitat Politècnica de Catalunya, Campus Nord B4, 08034 Barcelona, Spain}
\affiliation{Departament de Física de la Matèria Condensada, Universitat de Barcelona, Martí i Franquès 1, 08028 Barcelona, Spain}

\author{M. Carmen Miguel}
\affiliation{Departament de Física de la Matèria Condensada, Universitat de Barcelona, Martí i Franquès 1, 08028 Barcelona, Spain}
\affiliation{Institute of Complex System (UBICS), Universitat de Barcelona, Barcelona 08028, Spain}

\author{Romualdo Pastor-Satorras}
\affiliation{Departament de Física, Universitat Politècnica de Catalunya, Campus Nord B4, 08034 Barcelona, Spain}

\date{\today}

\begin{abstract}
 Moving animal groups transmit information through propagating waves or behavioral cascades, exhibiting characteristics akin to systems near a critical point from statistical physics. Using data from freely swimming schooling fish in an experimental tank, 
we investigate spontaneous behavioral cascades involving turning avalanches, where large directional shifts propagate across the group. We analyze several avalanche metrics and provide a detailed picture of the dynamics associated to turning avalanches, employing tools from avalanche behavior in condensed matter physics and seismology. Our results identify power-law distributions and robust scale-free behaviour through data collapses and scaling relationships, confirming a necessary condition for criticality in fish schools.
We explore the biological function of turning avalanches and link them to collective decision-making processes in selecting a new movement direction for the school. 
We report relevant boundary effects arising from interactions with the tank walls and  influential roles of boundary individuals. Finally, spatial and temporal correlations in avalanches are explored using the concept of aftershocks from seismology, revealing clustering of avalanche events below a designated timescale and an Omori law with a faster decay rate than observed in earthquakes.

\end{abstract}


  
\maketitle


\section{Introduction}

A fascinating and controversial hypothesis in biology is that some
systems may operate close to a critical point from statistical physics,
separating an ordered from a disordered state of the
system~\cite{stanleyIntroductionPhaseTransitions1987,binneyTheoryCriticalPhenomena1992,yeomansStatisticalMechanicsPhase1992}. Biological
systems at a critical point are believed to posses functional advantages such as
optimality in signal detection, storing and processing, large correlations in
coordinated behaviour and a wide spectrum of possible
responses~\cite{munozColloquiumCriticalityDynamical2018,
  moraAreBiologicalSystems2011, beggsBeingCriticalCriticality2012}. Criticality
is often associated to scale invariance, exemplified by power-law distributions
lacking relevant characteristic scales besides natural
cut-offs~\cite{binderFiniteSizeScaling1981,
  stanleyIntroductionPhaseTransitions1987,
  binneyTheoryCriticalPhenomena1992}. In particular, this is observed for
systems exhibiting spatiotemporal activity in the form of cascades or avalanches
with variable duration and size, which at the critical point are distributed as
power laws with anomalously large variance. There has been evidence of
criticality signatures in many different biological systems, ranging from neural
activity and brain networks, gene regulatory networks, collective behaviour of
cells or collective motion~\cite{munozColloquiumCriticalityDynamical2018,
  moraAreBiologicalSystems2011, romanczukPhaseTransitionsCriticality2022}.

The field of collective motion, in particular, studies the group movement
patterns exhibited by social organisms, such as flocks of birds, fish schools,
insect swarms, herds of mammals and human
crowds~\cite{sumpterCollectiveAnimalBehavior2010,vicsekCollectiveMotion2012}. In
this context, analytical and experimental studies of moving animal groups
suggest the existence of phase transitions between phases of coherent and
incoherent motion~\cite{tonerFlocksHerdsSchools1998, vicsekNovelTypePhase1995,
  couzinCollectiveMemorySpatial2002,   tunstromCollectiveStatesMultistability2013}. Moreover, groups of animals can
transmit information across the group in the form of propagating waves or
avalanches of behaviour, as occurs in fish schools
~\cite{brownFishCognitionBehavior2006, rosenthalRevealingHiddenNetworks2015,
  herbert-readInitiationSpreadEscape2015,
  lechevalSocialConformityPropagation2018,
  mugicaScalefreeBehavioralCascades2022, gomez-navaFishShoalsResemble2023},
honeybees~\cite{kastbergerSocialWavesGiant2008}, bird
flocks~\cite{pottsChoruslineHypothesisManoeuvre1984,
  procacciniPropagatingWavesStarling2011,attanasiInformationTransferBehavioural2014},
sheep herds~\cite{ginelliIntermittentCollectiveDynamics2015} or macaque
monkeys~\cite{danielsControlFiniteCritical2017}. Models of collective motion have also reproduced features of these phenomena~\cite{bialekStatisticalMechanicsNatural2012, cavagnaFlockingTurningNew2015, mwaffoAnalysisGroupFish2022, poelSubcriticalEscapeWaves2022}.  These behavioural cascades are
typically represented by behavioral shifts in the speed, acceleration or heading
of individuals, and can either arise spontaneously or from responses to
environmental cues, such as the presence of predators, food sources or obstacles. From a
biological point of view, they can occur when individuals follow the behaviour
of others without regarding their own
information~\cite{bikhchandaniTheoryFadsFashion1992}. From a physical
perspective, behavioral cascades can show signatures typical from systems
located near a critical point. Mainly, these signatures include large
susceptibility or sensitivity to
perturbations~\cite{attanasiFiniteSizeScalingWay2014,
  attanasiCollectiveBehaviourCollective2014, danielsControlFiniteCritical2017,
  poelSubcriticalEscapeWaves2022, gomez-navaFishShoalsResemble2023}, scale-free
correlations~\cite{cavagnaScalefreeCorrelationsStarling2010,
  attanasiCollectiveBehaviourCollective2014} and possible indications of
power-law behaviour in the avalanche size distribution~\cite{rosenthalRevealingHiddenNetworks2015,
  ginelliIntermittentCollectiveDynamics2015,
  mugicaScalefreeBehavioralCascades2022,
  gomez-navaFishShoalsResemble2023}. In addition, there is some evidence that
the state of criticality can be regulated by moving animal groups depending on
their needs~\cite{danielsControlFiniteCritical2017,
  sosnaIndividualCollectiveEncoding2019, poelSubcriticalEscapeWaves2022}, where
the avalanche dynamics may transition from being supercritical with local
changes propagating through the entire group, critical with changes propagating
at all possible scales of the system, or subcritical with changes remaining
local~\cite{pruessnerSelfOrganisedCriticalityTheory2012}.

In this study, we focus on analyzing the  properties of turning avalanches in freely moving fish~\cite{mugicaScalefreeBehavioralCascades2022}. These behavioral cascades involve the propagation of large changes in the heading direction of individuals within a group, often resulting in a reorientation of the group's global trajectory. Specifically, we examine spontaneous turning avalanches of schooling fish freely swimming in a tank. Our investigation unveils scale-free properties in the statistical distributions of different avalanche metrics and their dependence on the number of individuals in the school. Additionally, we investigate the origins of these avalanches, analyzing their triggering with respect to space, time, and individual initiators. Our findings reveal the relevance of interactions with tank walls 
and a larger influence of boundary individuals.  We also explore the dynamical evolution of
avalanches and its relation with the state of the school, as well as their spatial and temporal correlations. Within the limits of our experimental setup, our results strongly suggest the presence of a scale-free avalanche dynamics, that could be compatible with the school operating in the vicinity of a critical point.



\section{Avalanche definition and basic observables}

\begin{figure}[t]
\centering
  \subfloat[\label{fig:rate:turnRate}]{%
    \includegraphics[width=0.33\textwidth]{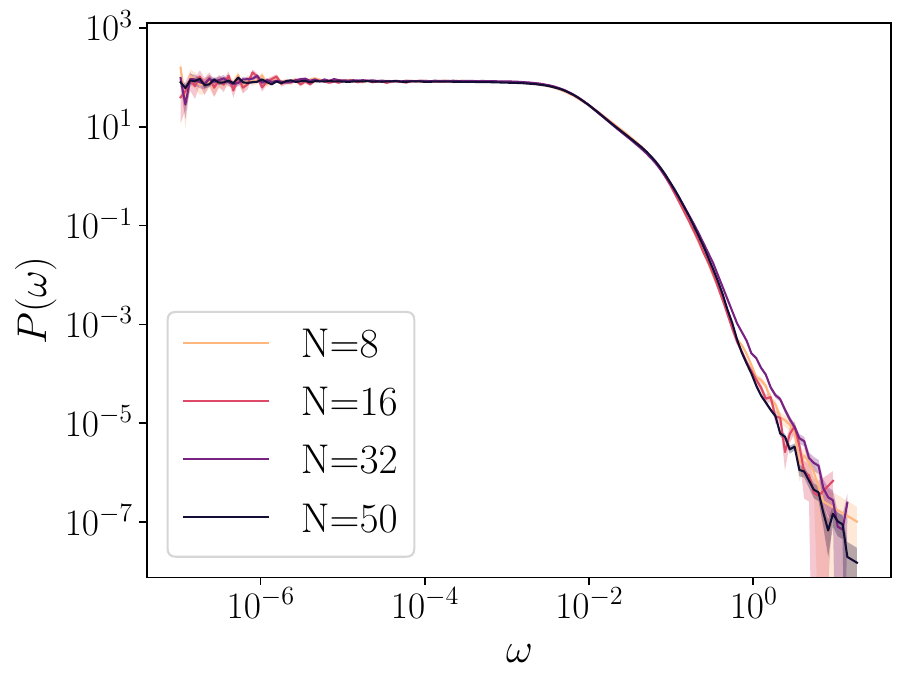}%
  }

  \subfloat[\label{fig:rate:avalancheRate}]{%
    \includegraphics[width=0.33\textwidth]{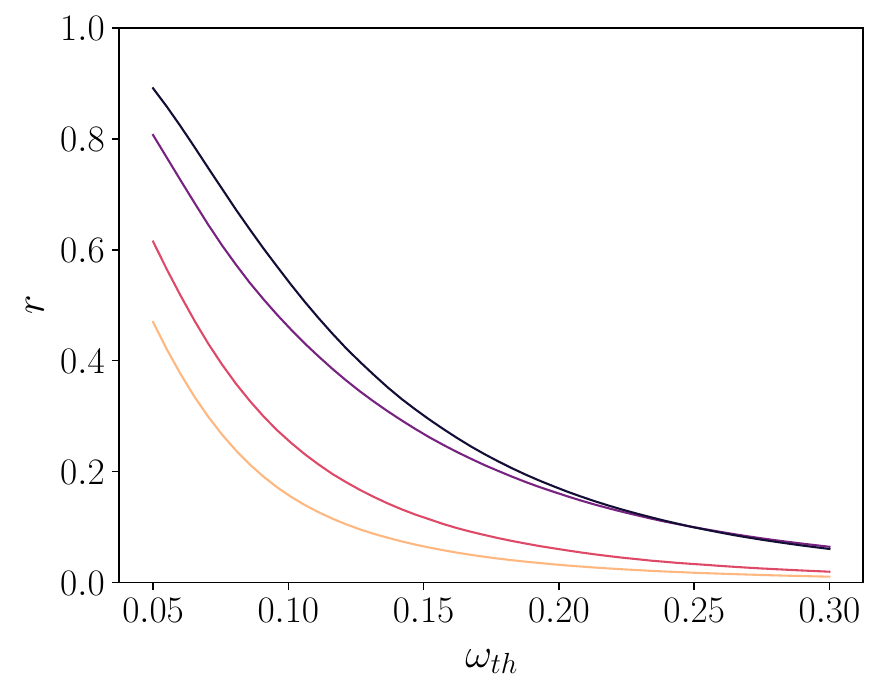}%
  }
  \caption{(a) PDF of the turning rate $\omega$ and (b) activity rate $r$ of
    turning avalanches as a function of the turning threshold $\omega_{th}$. The
    different curves correspond to experimental data from schools with different
    number of individuals $N$. Quantities are expressed in natural units of frames and pixels. \label{fig:rate}}
\end{figure}

Behavioral cascades in fish have been defined measuring changes of different
quantities. Here we focus on avalanches defined in terms of large changes in
the heading of individuals, given by their velocity
vector~\cite{mugicaScalefreeBehavioralCascades2022}. As experimental subject, we
consider the motion of $N=8, 16, 32$ and $50$ individuals of the species black
neon tetra \emph{Hyphessobrycon herbertaxelrodi}, a social fish that tends to
form polarized, compact and planar schools, freely swimming in an approximately
two-dimensional experimental tank. We recorded and digitized individual fish trajectories and calculated the corresponding velocities and accelerations 
(refer to Appendix~\ref{sec:methods:experimentalData} 
for  experimental and data acquisition details).  In
order to remove the dependency with the experimental frame rate of the
recordings, we measure the changes in time of the heading in terms of the
\emph{turning rate} $\omega$, defined as the absolute value of the angular
velocity, i.e.
\begin{equation}
  \omega  = \frac{\left| \vec{v} \times \vec{a} \right|}{v^2},
  \label{eq:6}
\end{equation}
where $\vec{v}$ and $\vec{a}$ are the instantaneous velocity and acceleration of
an individual respectively, and $v$ is the modulus of the instantaneous
velocity.  See Appendix~\ref{sec:methods:turning rate formula} 
for a
derivation of this expression. We consider the absolute value due to symmetry in
the turning direction. 

In Fig.~\ref{fig:rate:turnRate} we show the probability density function (PDF)
of the turning rate $P(\omega)$, for schools of different number of individuals
$N$. Here and in the following, we work in natural units of pixels and frames
for distance and time, respectively. In addition, error bands in the PDF plots
are calculated from the standard deviation of a Bernoulli distribution with the
probability given by the fraction of counts in each bin of the numerical
PDF~\cite{salvatPENELOPE2018CodeSystem2019}. As we can see, schools of different
number of individuals show essentially the same behavior in their turning rate
distributions. Most of the time, the turning rate is very small and uniformly
distributed, corresponding to fish swimming locally in a straight trajectory.
In some instances, however, large turning rates can be observed, in which
individuals swiftly rearrange their headings and thus reorient their direction
of motion.

Inspired by avalanche behavior in condensed matter
physics~\cite{zapperiCracklingNoiseStatistical2022}, we define avalanches by
introducing a \emph{turning threshold} $\omega_{th}$ separating small from large
turns~\cite{mugicaScalefreeBehavioralCascades2022}. Considering an
\textit{active} fish as one with a turning rate $\omega > \omega_{th}$, we
introduce the dynamical variable $n_t$ defined as the number of active fish
observed at frame $t$. Then, sequences of consecutive frames in which $n_t >0$
(i.e. in which there is at least one active fish) define a \emph{turning
  avalanche}. In the Supplementary Video~\ref{app:video:1}~\footnote{See Supplemental Material at
  \url{http://link.aps.org/supplemental/XXX} for supplemental videos and figures.} we show some examples
of large turning avalanches for a school of $N=50$
fish.

The most basic characterization of turning avalanches is given by the duration
$T$ and size $S$ of avalanches, and by their inter-event time $t_i$. An
avalanche starting at frame $t_0$ has \emph{duration} $T$ if the sequence of
dynamic variables $n_t$ fulfills $n_{t_0-1} = 0$, $n_t > 0$ for
$t = t_0, \ldots, t_0 + T-1$, and $n_{t_0 + T} = 0 $.  The \emph{size} $S$ of an
avalanche is given by the total number of active fish in the whole duration of
the avalanche, i.e. $S = \sum_{t = t_0}^{t_0 + T -1} n_t$. The \emph{inter-event
  time} $t_i$ between two consecutive avalanches is given by the number of
frames between the end of one avalanche and the start of the next one, that is,
by a sequence fulfilling $n_{t_f} > 0$, $n_t = 0$ for
$t = t_f+1, \ldots, t_f + t_i$, and $n_{t_f + t_i + 1} > 0$, where $t_f$
indicates the last frame of the first
avalanche~\cite{zhongBurstinessInformationSpreading2023}.

The effects of the turning threshold in avalanches can be assessed with the
\emph{activity rate} $r$, defined as the probability that a randomly chosen
frame belongs to an avalanche. We compute it as the ratio between the number of
frames with activity $n_t > 0$ and the total number of frames in the
experimental series. As we can see from Fig.~\ref{fig:rate:avalancheRate}, for
fixed $N$ the activity rate decreases with the turning threshold $\omega_{th}$,
since by increasing $\omega_{th}$ we are decreasing the turning rates that we
consider large and we find less frames with $n_t > 0$. On the other hand,
increasing the number of individuals $N$ at fixed $\omega_{th}$ results in an
increase of the activity rate. We can interpret this as a school with larger
number of individuals has a higher probability for any of them to display a
large turning rate.

Realistic values of $\omega_{th}$ used to compute avalanches are estimated to lie within the range $\omega_{th} \in [0.01, 0.3]$. Smaller values result in infinite avalanches that span the entire duration of the experiment, whereas larger values produce very few avalanches.

\section{Statistical distributions}
\label{sec:stat-distr}

\begin{figure}[t]
\centering
\subfloat[\label{fig:measures:duration}]{%
  \includegraphics[width=0.23\textwidth]{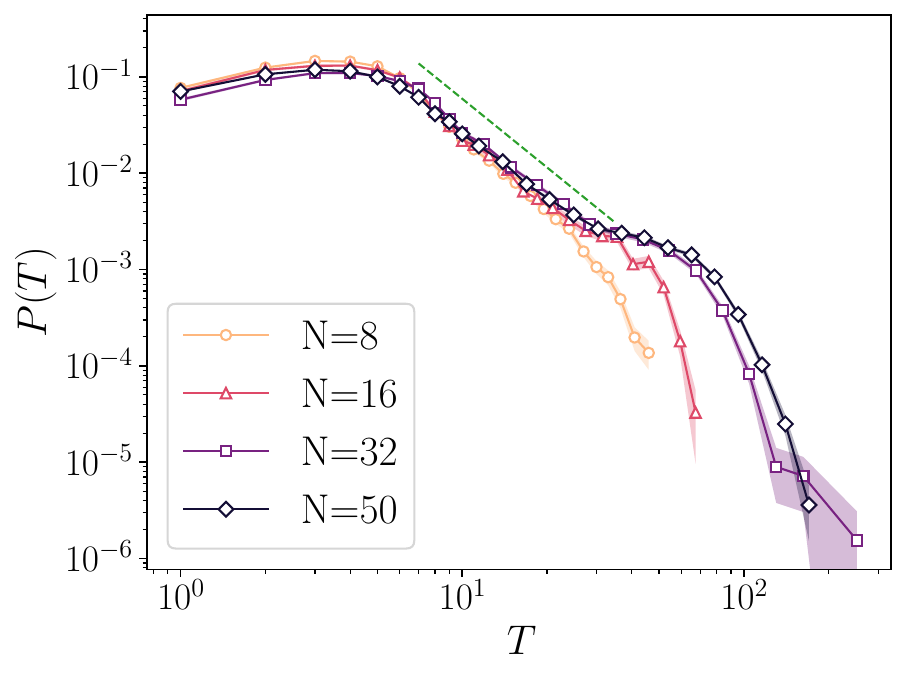}%
}
\hspace{0.01\textwidth}
\subfloat[\label{fig:measures:size}]{%
  \includegraphics[width=0.23\textwidth]{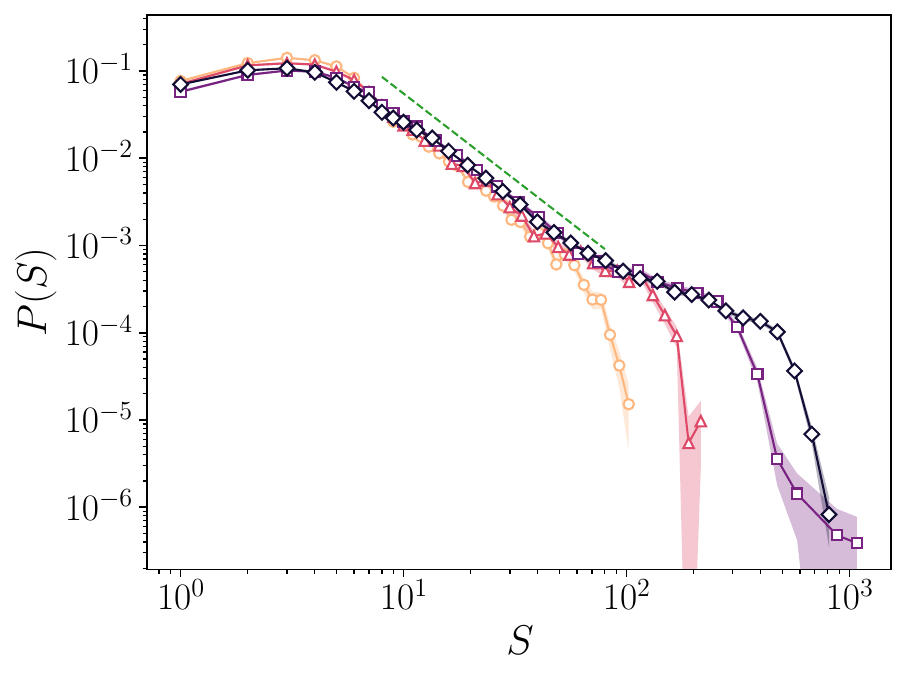}%
}

\subfloat[\label{fig:measures:T-S}]{%
  \includegraphics[width=0.23\textwidth]{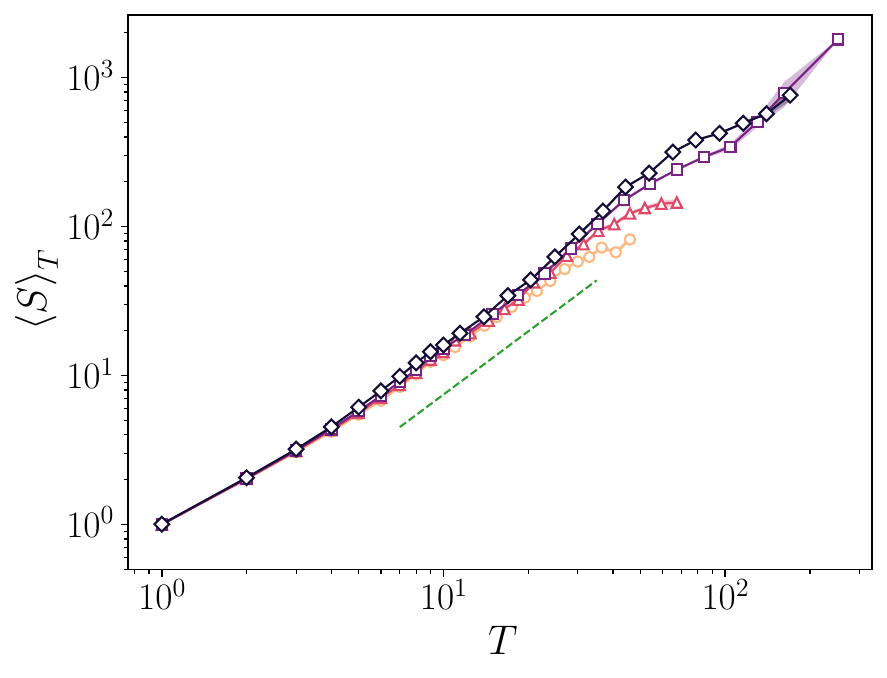}%
}
\hspace{0.01\textwidth}
\subfloat[\label{fig:measures:interEvent}]{%
  \includegraphics[width=0.23\textwidth]{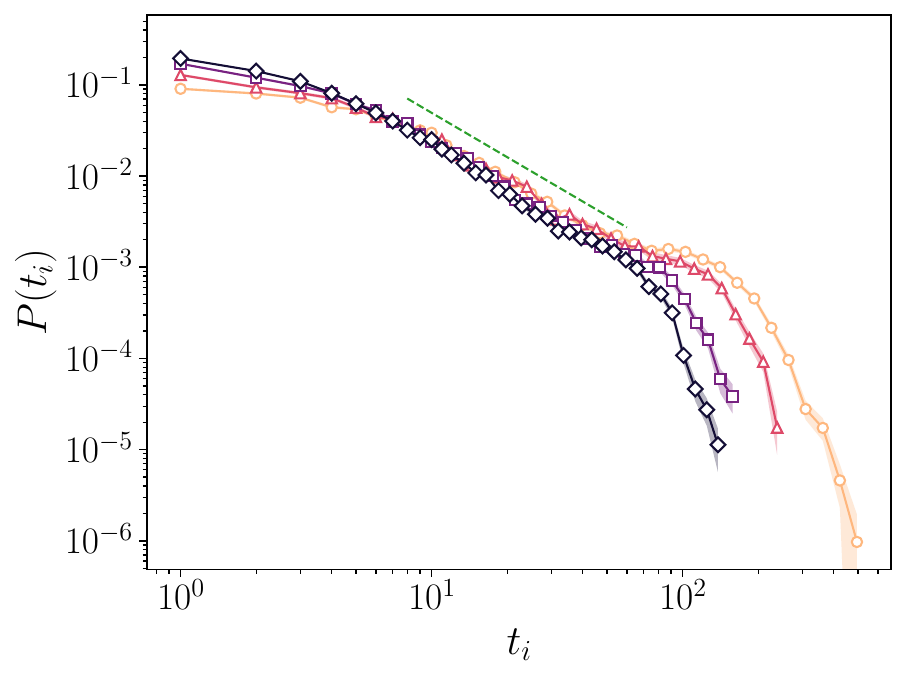}%
}
\caption{(a) PDF of the duration $T$, (b) PDF of the size $S$, (c) average size
  $\left < S\right >_T$ as a function of the duration $T$ and (d) PDF of the
  inter-event time $t_i$ for $\omega_{th}=0.1$. The different curves correspond
  to schools of different number of individuals $N$. The exponents from the
  green dashed power laws are (a) $\alpha = 2.4 \pm 0.2$, (b)
  $\tau=1.97 \pm 0.14$, (c) $m=1.41\pm 0.06$ and (d)
  $\gamma = 1.62\pm 0.08$.}\label{fig:measures}
\end{figure}

In Figs.~\ref{fig:measures:duration} and~\ref{fig:measures:size} we show the
distributions of the duration $T$ and size $S$, respectively, obtained for a fixed
turning threshold $\omega_{th}=0.1$ and for schools of different number of
individuals $N$.  We find that both PDFs show a power-law scaling region of the
form
\begin{equation}
  \label{eq:1}
  P(T) \sim T^{-\alpha}, \qquad  P(S) \sim S^{-\tau},
\end{equation}
limited by a peak for low values and a shoulder or bump with a fast decaying (exponential) tail for high values. The characteristic exponents $\alpha$ and $\tau$, obtained from a linear
regression in double logarithmic scale in the scaling region, take the values
$\alpha=2.4 \pm  0.2$ and $\tau=1.97 \pm 0.14$, where the error bars represent $95\%$ confidence intervals. Different values of $\omega_{th}$
lead to similar average exponents (e.g $\alpha = 2.9 \pm 0.8$ and
$\tau = 2.4\pm0.4$ for $\omega_{th} = 0.15$, see Supplementary
Fig.~\ref{app:fig:measures}). These exponents align with previous estimates derived from smaller statistics and using a different definition of turning fish~\cite{mugicaScalefreeBehavioralCascades2022}.
Interestingly, distributions for schools of
different number of individuals collapse onto the same functional form with the
exception of the tail, which can be interpreted in terms of finite
size effects, as larger schools tend to create avalanches of larger duration and
size. 

The duration and size of individual avalanches are not independent, as we can
check by plotting the average size $\left < S \right >_T$ of avalanches of
duration $T$, see Fig.~\ref{fig:measures:T-S}. From this figure we can observe a
superlinear behavior
\begin{equation}
\left < S \right >_T \sim T^m,\label{eq:averageSize_T}
\end{equation}
with $m=1.41\pm0.06$. The value of $m$ can be related to the exponents of the
duration and size distributions
as~\cite{pruessnerSelfOrganisedCriticalityTheory2012,mugicaScalefreeBehavioralCascades2022}
\begin{equation*}
  m = \frac{\alpha - 1}{\tau - 1}.
\end{equation*}
Our experimental value $m$ is fully compatible with the theoretical prediction
$m = 1.4 \pm 0.3$ for $\omega_{th} = 0.1$ (experimental $m=1.35\pm0.16$ and
theoretical prediction $m = 1.4\pm0.7$ for $\omega_{th} = 0.15$, see
Supplementary Fig.~\ref{app:fig:measures:T-S}).

In Fig.~\ref{fig:measures:interEvent} we show the PDF of the inter-event time
$t_i$ for $\omega_{th}=0.1$ and for schools of different number of individuals
$N$. We find again an intermediate scale-free region, limited between the small
time behavior and a shoulder with an exponentially decreasing tail. Here also plots for
different number of individuals $N$ collapse on the same functional form, with
the exception of the tail. A fit to the form
\begin{equation*}
  \label{eq:2}
  P(t_i)\sim t_i^{-\gamma}
\end{equation*}
in the scaling region leads to an average exponent $\gamma = 1.62\pm0.08$ ($\gamma = 1.63\pm0.04$ for $\omega_{th} = 0.15$, see Supplementary Fig.~\ref{app:fig:measures:interEvent}). It is noteworthy
that the behavior of the decaying tails with $N$ is reversed with respect to the
duration and size PDFs, with larger number of individuals leading to smaller
inter-event times.  This observation is consistent with the behavior of the activity
rate $r$, as schools with larger number of individuals have a higher probability
to be in an avalanche.

\begin{figure}[t]
\centering
\subfloat[\label{fig:measuresFixedR:duration}]{%
  \includegraphics[width=0.23\textwidth]{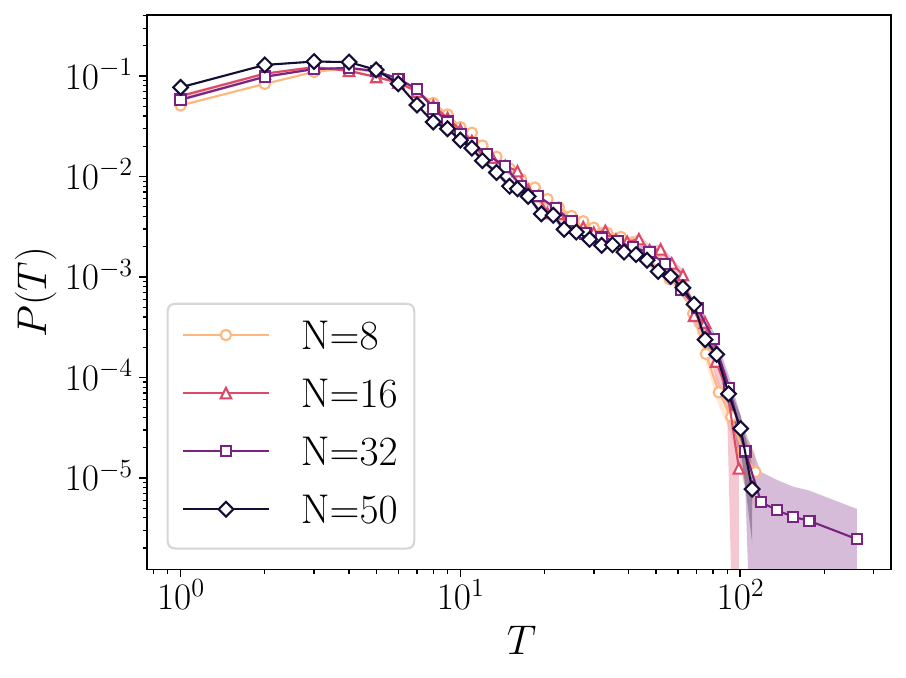}%
}
\hspace{0.01\textwidth}
\subfloat[\label{fig:measuresFixedR:size}]{%
  \includegraphics[width=0.23\textwidth]{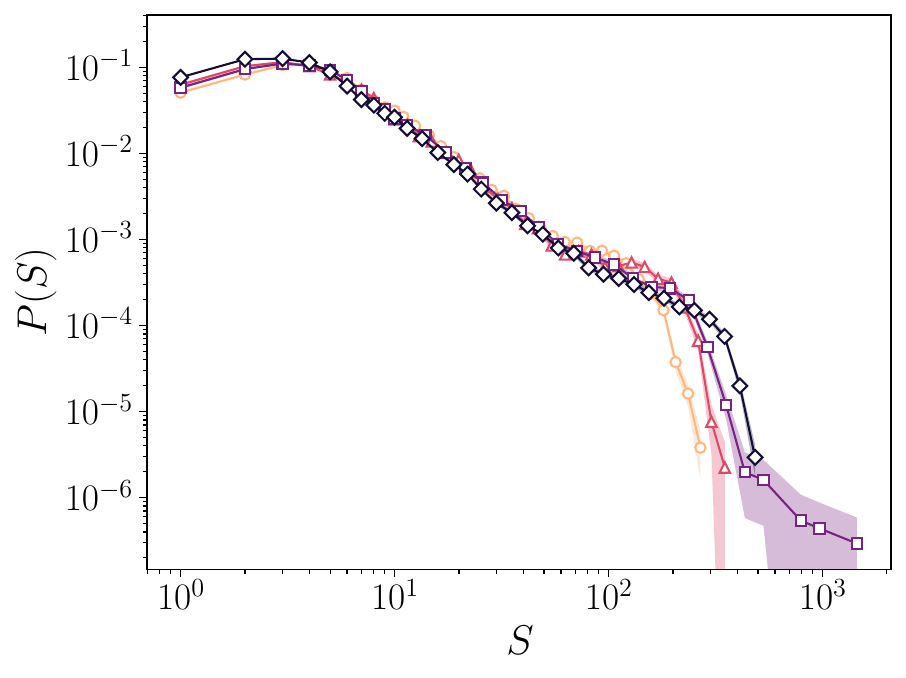}%
}

\subfloat[\label{fig:measuresFixedR:tInterEvent}]{%
  \includegraphics[width=0.24\textwidth]{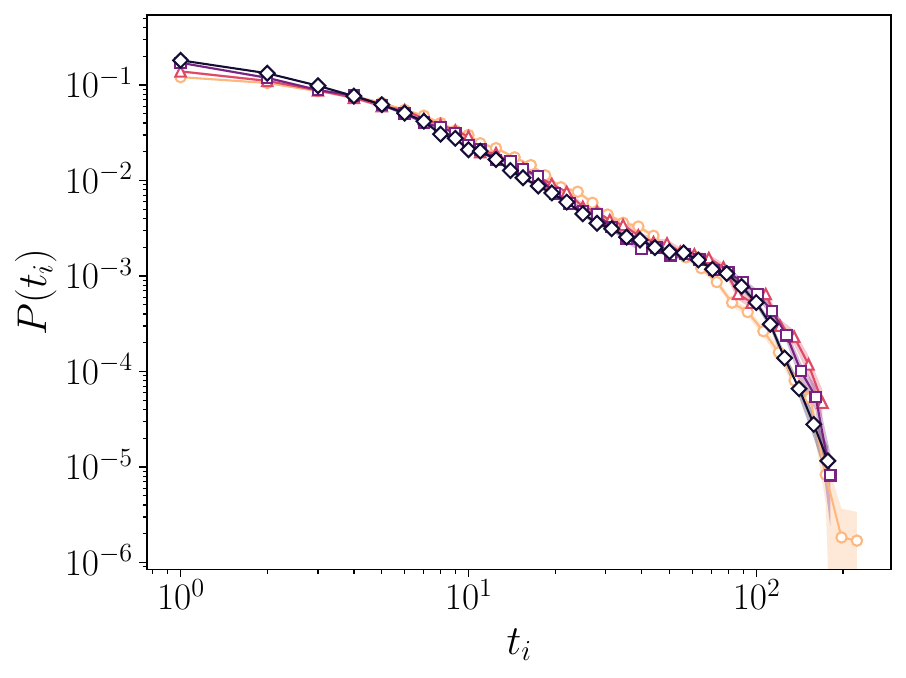}%
}
\subfloat[\label{fig:measuresFixedR:tInterEvent-collapse}]{%
  \includegraphics[width=0.24\textwidth]{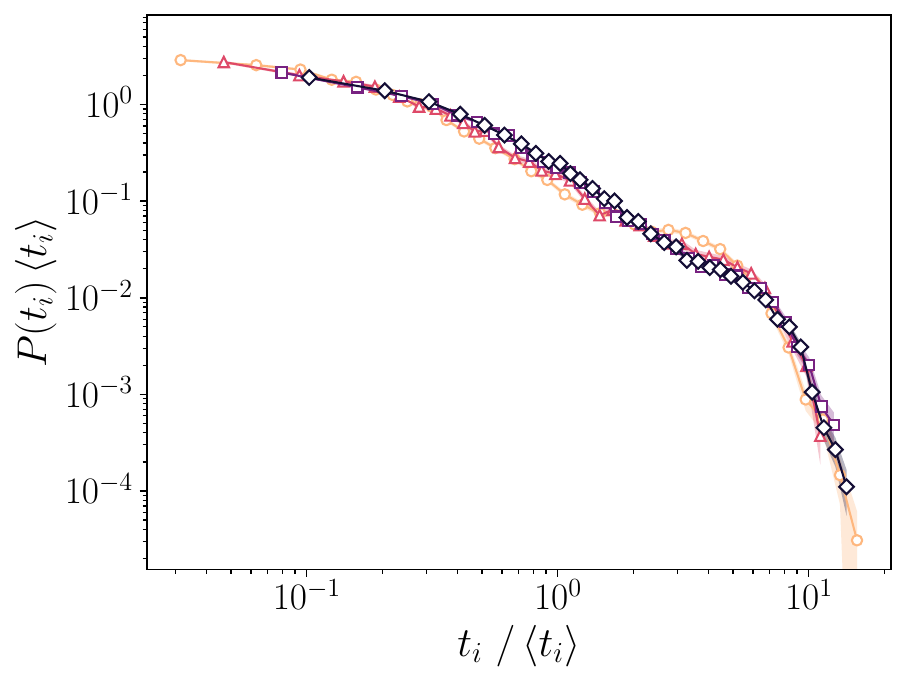}%
}
\caption{Data collapse for the PDFs of (a) the duration $T$, (b) the size $S$
  and (c) the inter-event time $t_i$ for schools of different number of
  individuals $N$ considering avalanches with a fixed activity rate $r=0.4$
  (corresponding to $\omega_{th} = 0.055, 0.076 , 0.11, 0.13$ for
  $N=8, 16, 32, 50$ respectively).  (d) Data collapse of the inter-event time
  given by Eq.~\eqref{eq:dataCollapse_interevent} for
  $\omega_{th}=0.1$. }\label{fig:measuresFixedR}
\end{figure}

\section{Data collapse}
\label{sec:data-collapse}

The dependency of the tails in the duration and size distributions
with the school size $N$ observed above, and with the turning threshold $\omega_{th}$ reported
in~\cite{mugicaScalefreeBehavioralCascades2022}, suggests the possibility of a
relationship between $\omega_{th}$ and $N$ resulting in avalanches with
collapsing distributions. In order to test for this hypothesis, we select the
threshold $\omega_{th}$ that, for each value of $N$, leads to a fixed activity
rate $r=r_0$. From Fig.~\ref{fig:rate:avalancheRate} we estimate, for
$r_0 = 0.4$, $\omega_{th} = 0.055, 0.076 , 0.11, 0.13$ for $N=8, 16, 32, 50$,
respectively. We plot the resulting distributions in
Figs.~\ref{fig:measuresFixedR:duration},~\ref{fig:measuresFixedR:size}
and~\ref{fig:measuresFixedR:tInterEvent} for the duration $T$, size $S$ and
inter-event time $t_i$, respectively.  In a system with no temporal correlations
in the activity of individuals, a fixed activity rate results in duration and
inter-event time distributions collapsing onto the same functional, exponential
forms, see Appendix~\ref{sec:methods:null_model}. Surprisingly, even if
this is not the case for empirical turning avalanches in schooling fish, both
the duration and inter-event time distributions achieve a data collapse at fixed
$r$.  On the other hand, the size distributions do not collapse perfectly,
possibly because of correlations in the turning rates of individuals at a given
frame, which results in more active individuals in an avalanche frame for
schools of larger number of individuals. Interestingly, also in the uncorrelated
case the size distributions are not expected to collapse, see Appendix~\ref{sec:methods:null_model}.

On a similar note, for avalanches of self-organized critical phenomena across different contexts, it has been found that the inter-event time distributions
can be collapsed into the scaling form~\cite{corralUniversalLocalUnified2004,
  baroAvalancheCorrelationsMartensitic2014}
\begin{equation}
  P(t_i) = \frac{1}{\left< t_i \right>} \Phi \left( \frac{t_i}{\left< t_i
      \right>}\right),
  \label{eq:dataCollapse_interevent}
\end{equation}
where $\Phi(x)$ is a universal scaling function, and the only characteristic
scale is the average inter-event time $\left< t_i \right>$. In
Fig.~\ref{fig:measuresFixedR:tInterEvent-collapse} we show this sort of
collapse for a turning threshold  $\omega_{th} = 0.1$; as we can see,
it also applies to turning avalanches in schooling fish. This reveals
self-similar behaviour, with the inter-event time
distributions only differing in their average value for schools of different
number of individuals. In the uncorrelated case, this collapse is also
recovered, but now only in the limit of a large average inter-event time, see Appendix~\ref{sec:methods:null_model}.

\begin{figure}[t]
\centering
\includegraphics[width=0.33\textwidth]{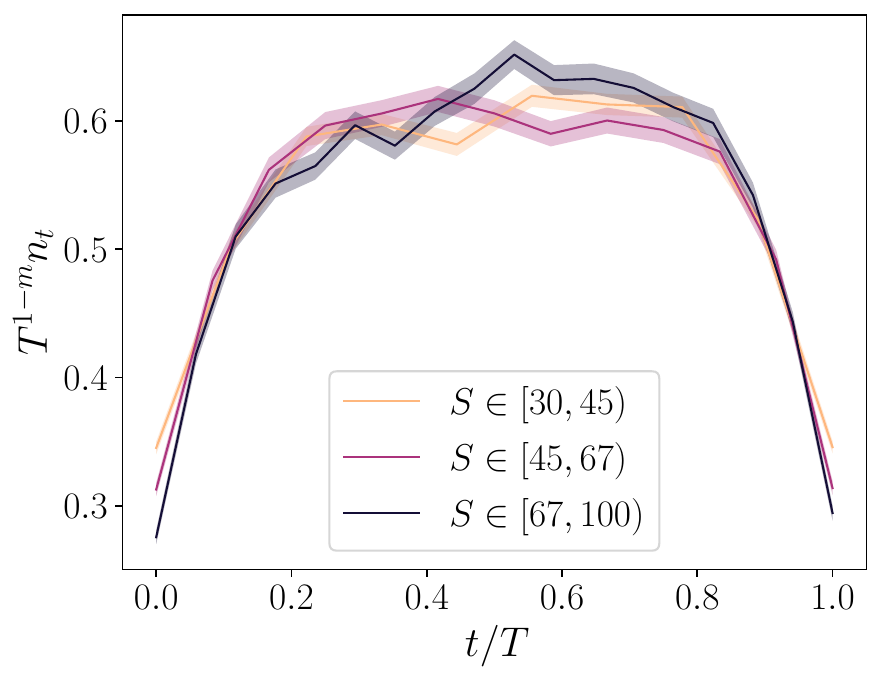}%
\caption{Rescaled avalanche shape $T^{1-m} n_t$ as a function of the normalized
  time $t/T$. Avalanche shapes are averaged over similar sizes $S$ within the
  power-law scaling region of the size distribution.}\label{fig:shape}
\end{figure}

As a final check of the scale-free nature of turning avalanches, we consider the
scaling of the \emph{avalanche shape} $n_t$, defined by the number of active
individuals of a turning avalanche at the frame $t$ of its
duration~\cite{kuntzNoiseDisorderedSystems2000,kuntzNoiseDisorderedSystems2000}. Many
scale-free avalanche systems exhibit a collapse behavior in the avalanche shape
given by the scaling relation
\begin{equation*}
    n_t = T^{m-1} \Phi (t/T),
    \label{eq:sethna}
\end{equation*}
where $m = (\alpha-1)/(\tau-1)$ is the exponent relating the average avalanche
size $\left< S\right>_T$ with the duration $T$,
Eq.~\eqref{eq:averageSize_T}, and $\Phi(z)$ is a universal scaling
function~\cite{sethnaCracklingNoise2001,
  beggsBeingCriticalCriticality2012, kuntzNoiseDisorderedSystems2000,
  friedmanUniversalCriticalDynamics2012}. In the case of turning avalanches, this scaling behavior is recovered
in avalanches within the power-law scaling regime of the size distribution, as
shown in Fig.~\ref{fig:shape}. In this plot, the avalanche shape is computed
normalizing the avalanche time frame $t$ by its duration $T$ and averaging over
avalanches in a given size range.  We use the value $m=1.41$ obtained in the numerical
analysis of the duration and size distributions.

\section{Avalanche triggering}\label{sec:avalanche triggering}

\begin{figure}[t]
\centering
\subfloat[\label{fig:spatiotemporal loc:norm}]{%
  \includegraphics[width=0.22\textwidth]{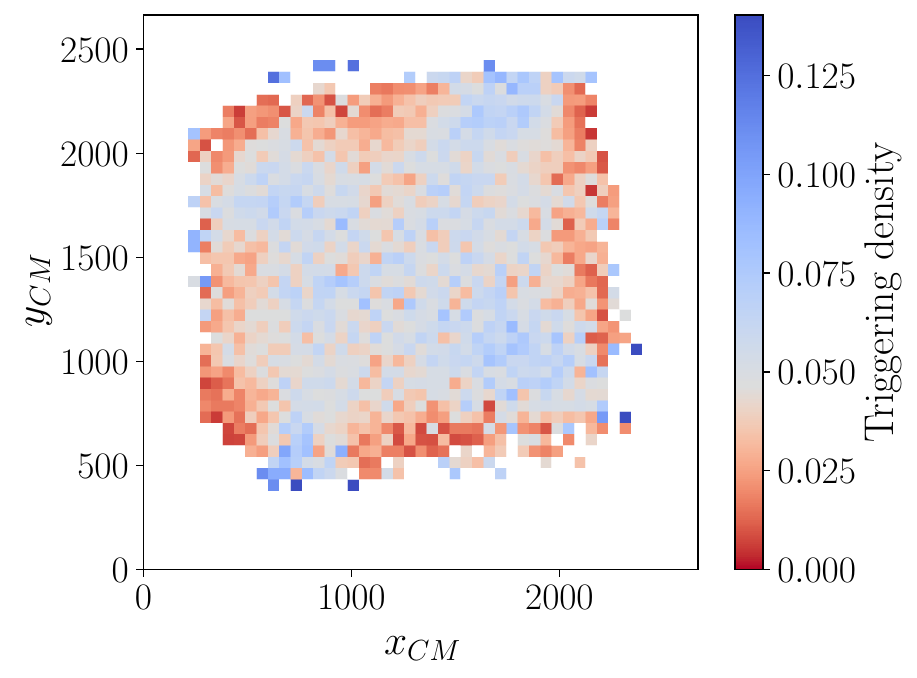}%
}
\hspace{0.01\textwidth}
\subfloat[\label{fig:spatiotemporal loc:size}]{%
  \includegraphics[width=0.22\textwidth]{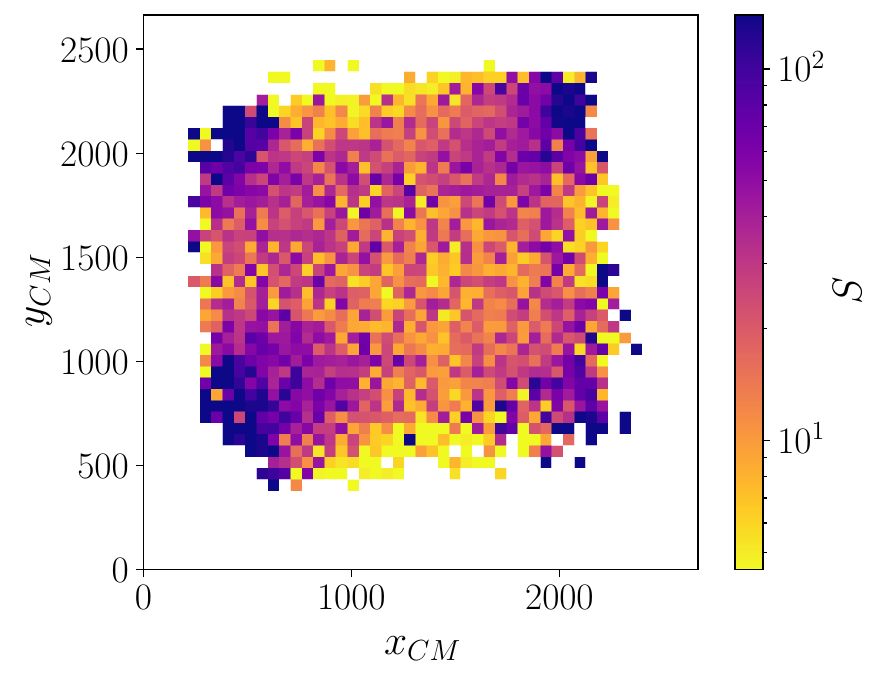}%
}

\subfloat[\label{fig:spatiotemporal loc:vCM_t}]{%
  \includegraphics[width=0.48\textwidth]{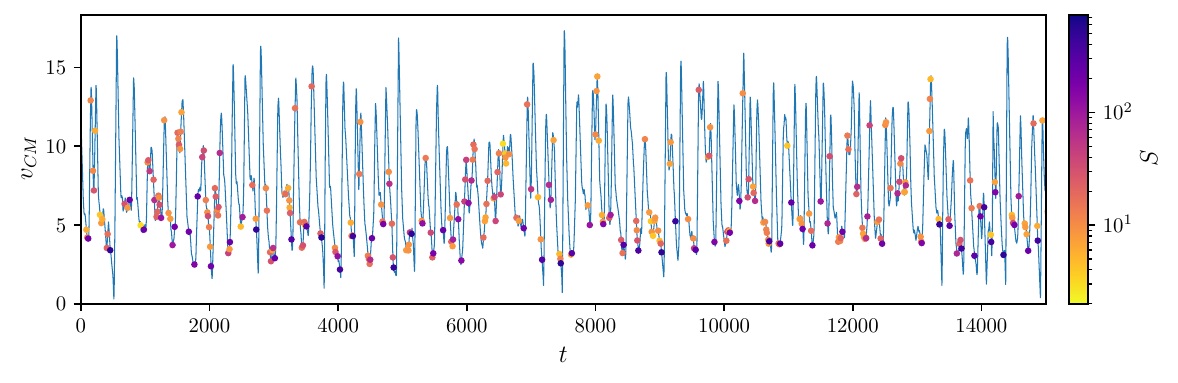}
}

\subfloat[\label{fig:initiators:lab}]{%
  \includegraphics[width=0.22\textwidth]{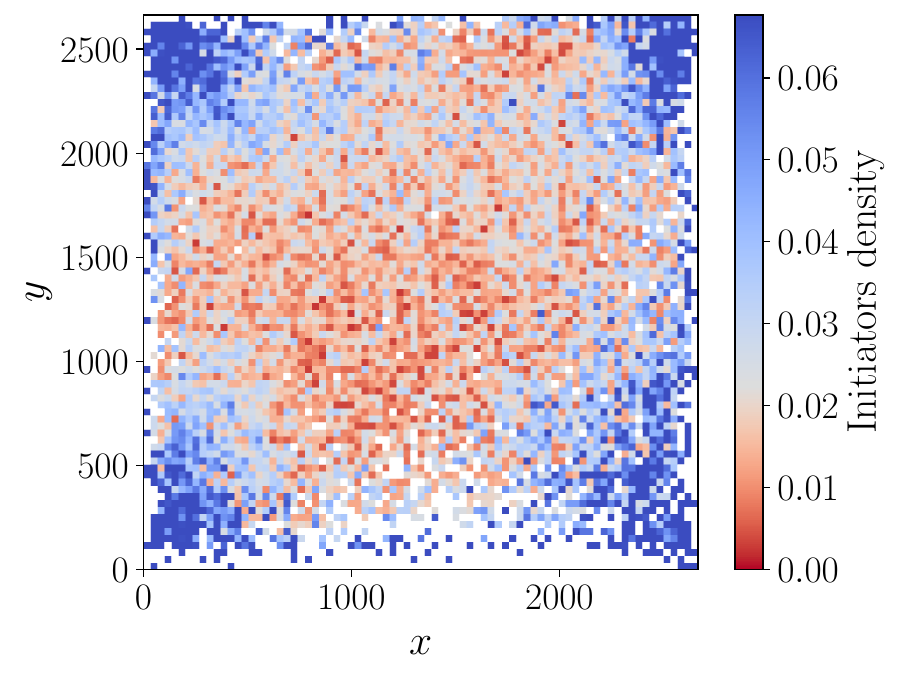}%
}
\hspace{0.01\textwidth}
\subfloat[\label{fig:initiators:CM}]{%
  \includegraphics[width=0.22\textwidth]{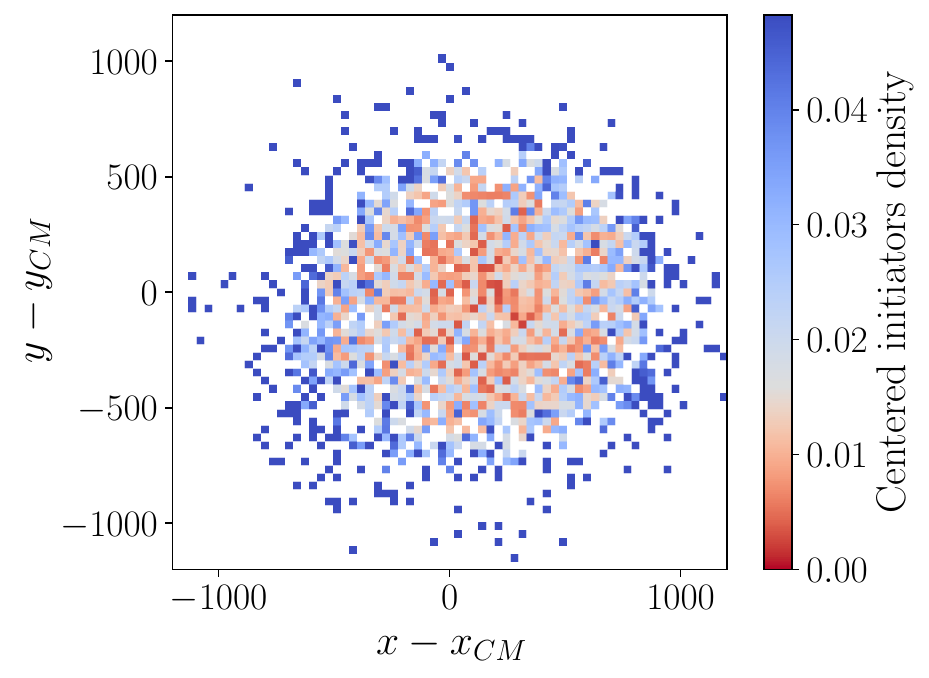}%
}
\caption{Avalanche triggering in space, time and within the group. (a) Density
  for the position of the center of mass (CM) $\vec{x}_{CM}$ at the start $t_0$
  of an avalanche (the triggering location) normalized against all trajectories
  of the center of mass, (b) average size $S$ for triggering locations of
  avalanches, (c) in blue the temporal evolution for the center of mass speed
  $v_{CM}$ and in dots avalanches triggered at the given speed $v_{CM}$ and time
  $t_0$ and coloured by their size $S$, (d)-(e) density for the position of
  initiators normalized against the positions of all individuals at the start
  $t_0$ of an avalanche for (d) the laboratory reference frame and (e) the
  center of mass reference frame and only for centered individuals. In (a), (d)
  and (e) the grey colour in the colormap corresponds to the expected density in
  the absence of correlations, given by the total counts of the quantity
  considered divided by the total counts of the normalization. In (c) we only
  plot avalanches that propagated to individuals other than the ones active in
  the first frame of the avalanche.  In (e) the $y$-coordinate is oriented along
  the direction of motion of the group given by the center of mass velocity.}
\label{fig:spatiotemporal loc}
\end{figure}

In this section we explore whether avalanches are triggered in some preferential
points in space or time, as well as by particular individuals in the group. Here
and in the following sections we show results for avalanches in a school of
$N=50$ individuals, which have the longest recording time, and a turning threshold $\omega_{th} = 0.1$.

A plausible hypothesis is that avalanches are more frequently triggered near the
tank walls due to boundary effects. These could arise when fish are approaching
a wall and need to perform a large turn in order to avoid colliding with it. To
check this hypothesis we consider the position for the center of mass (CM)
$\vec{x}_{CM}$ of the school, defined as
\begin{equation*}
  \vec{x}_{CM} \equiv \frac{1}{N} \sum_i \vec{x}_i,
\end{equation*}
where $\vec{x}_i$ are the positions of the fish at a given instant of time.  We
define the \emph{triggering location} of an avalanche as the position of the CM
at the first frame $t_0$ of the avalanche. We study the distribution of
triggering locations on the surface of the tank.  Because fish do not swim
uniformly all around the tank, in order to extract a statistically significant
density of triggering locations we normalize their counts against the counts of
all observed positions of CM along the time evolution of the school. We show
this in Fig.~\ref{fig:spatiotemporal loc:norm}, where the axis orientations
correspond to the tank walls. The grey region in the colormap, separating the
low density (red) and high density (blue) values, corresponds to the expected
density in the absence of correlations, which we calculate from the total counts
of triggering locations divided by the total counts of positions of CM. As we
can see in this plot, the distribution of avalanches in the tank is quite
homogeneous, although there is a slight tendency for avalanches to occur away
from the walls.  However, if we display the average size $S$ of avalanches
generated at the different triggering locations, we obtain a different picture,
Fig.~\ref{fig:spatiotemporal loc:size}, in which avalanches of larger sizes tend
to occur more frequently near the tank corners. This observation suggests that
interactions with the tank walls indeed promote the emergence of large turning
avalanches, resulting in important orientation rearrangements of the school.


Since large avalanches seem to be originating from interactions with the walls,
we investigate whether these interactions are responsible for the
shoulder or bump observed in the the tails of the duration and size
  distributions. They are particularly noticeable in groups with larger numbers of individuals ($N=32$ and $N=50$), which are expected to have more frequent interactions with the tank walls. This feature, known as \emph{dragon kings}, breaks the
  power-law paradigm by displaying overrepresented extreme
  events~\cite{sornetteDragonKingsBlackSwans2009,
    sornetteDragonkingsMechanismsStatistical2012,
    mikaberidzeDragonKingsSelforganized2023}. Dragon kings are typically
  generated by different mechanisms than smaller events, which, in this case,
  may be wall interactions. To explore this, we analyze the statistical
distributions of avalanches with triggering locations away from the walls, which
we restrict to occur inside the square positioned at the center of the tank with
side $L/3$, where $L$ is the side of the tank (refer to Supplementary
Fig.~\ref{app:fig:centralHist}). We expect these avalanches to  arise spontaneously and not be promoted by interactions with the tank walls. Despite limited statistics, dragon
  kings are no longer observed, and the distributions now showcase extended
  power-law regions with the same characteristic exponents as previously
  measured. We quantified the presence of dragon kings in the size distribution
  for $N=50$ with a statistical dragon kings detection
  test~\cite{mikaberidzeDragonKingsSelforganized2023,
    janczuraBlackSwansDragonkings2012} (see Appendix~\ref{sec:methods:statistical_dragon_kings_test}). Employing a significance level
  $\alpha=0.05$ for the null hypothesis that there are no dragon kings, the test
  confirms dragon kings ($p$-value $p < 10^{-15}$) for the total size distribution
  (Fig.~\ref{fig:measures:size}); and rejects their presence ($p=0.1$) for the
  size distribution restricted to the central region of the tank (Supplementary
  Fig.~\ref{app:fig:centralHist}).

To understand temporal triggerings of avalanches, we study how the avalanche
starting time $t_0$ relates to the group dynamics represented by the
\emph{center of mass speed}~$v_{CM}$, which is defined as
\begin{equation}
  v_{CM} \equiv \left | \frac{1}{N} \sum_i \vec{v}_i \right |.
  \label{eq:3}
\end{equation}
The center of mass speed is characterised for having oscillations due to a
burst-and-coast mechanism of the
individuals~\cite{weihsEnergeticAdvantagesBurst1974,
  harpazDiscreteModesSocial2017, liBurstandcoastSwimmersOptimize2021}, with
increases associated to an active phase powered by the fish muscles and
decreases coming from a passive gliding phase. In Fig.~\ref{fig:spatiotemporal
  loc:vCM_t} we plot, for a time window of 5~min from a single recording, the temporal evolution of the
center of mass speed as the blue line. We mark with dots avalanches triggered at
the corresponding time $t_0$ and speed $v_{CM}$, color-coded by their size
$S$. We only consider avalanches that propagated to individuals other than the
ones active in the first frame of the avalanche. As we can observe, while small
size avalanches tend to be randomly distributed over different
values of $v_{CM}$, large avalanches are more often located near the minima of
the speed, even when the minimum changes across time. We notice that this
behavior does not originate from small speeds being related to large turning
rates, because we find the turning rate is inversely related to the speed only
for $v_{CM} < 4$ and appears to be independent for larger speeds (see
Supplementary Fig.~\ref{app:fig:omega-v}). Instead, this suggests that
large avalanches may emerge from turnings related to decision-making processes
occurring at the onset of the active phase of the burst-and-coast
mechanism~\cite{herbert-readHowPredationShapes2017,
  harpazDiscreteModesSocial2017, caloviDisentanglingModelingInteractions2018}.

Apart from the spatiotemporal triggering of avalanches, we can study how
avalanches are triggered at the individual level within the school considering
avalanche \emph{initiators}, defined as the individuals that are active on the
first frame of the avalanche. Previously it was observed that some individuals
have a probability larger than random fluctuations to be the initiators of
behavioral cascades~\cite{mugicaScalefreeBehavioralCascades2022}.  Here instead
we focus on the location of individual initiators within the experimental tank
and inside the school. Again, we have to keep in mind that individuals are not
located uniformly around the tank at the start of an avalanche. Therefore, in
order to extract a statistically significant density of initiators locations
within the group, we normalize their counts against the counts of the positions
of all individuals at the onset time $t_0$ of the avalanche. We show the
resulting plot in Fig.~\ref{fig:initiators:lab}. We find that initiators tend to
accumulate near the tank walls, and particularly at the corners. This is
compatible with the idea that large turning avalanches are promoted by
interactions with the tank walls.


In order to explore the natural relative position of avalanche initiators within
the school, we select individuals that do not have relevant interactions with
the tank walls. We define \emph{centered individuals} as those that are
positioned in the central square of the tank with side $L/3$, where $L$ is the
side of the tank. If we plot the density of the positions of centered initiators
within the tank normalised by the positions of all centered individuals at the
onset time $t_0$ of an avalanche (see Supplementary
Fig.~\ref{app:fig:lab_centred}), indeed we see a uniform pattern that confirms
the idea that centered initiators do not experience significant interactions
with the tank walls. We study the relative position of centered initiators
within the school in Fig.~\ref{fig:initiators:CM}, where we plot the density of
the positions of centered initiators normalized against all centered individuals
at the triggering time $t_0$ of the avalanche in the center of mass reference
frame. In this plot the $y$-coordinate is directed along the direction of motion
of the center of mass. As we can see, initiators of avalanches away from the
tank walls accumulate on the boundary of the school and without any preferred
direction along the movement of the group.

\section{Dynamical evolution of avalanches}

\begin{figure*}[t!p]
  \subfloat[\label{fig:dynamics:vCM}]{%
  \includegraphics[height=0.22\textwidth]{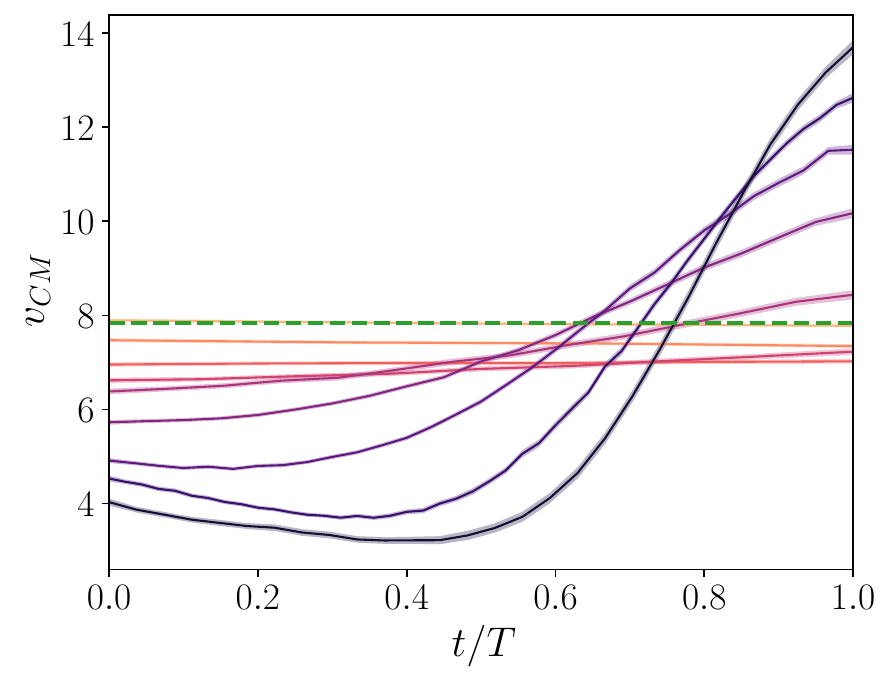}%
}
\subfloat[\label{fig:dynamics:phi}]{%
  \includegraphics[height=0.22\textwidth]{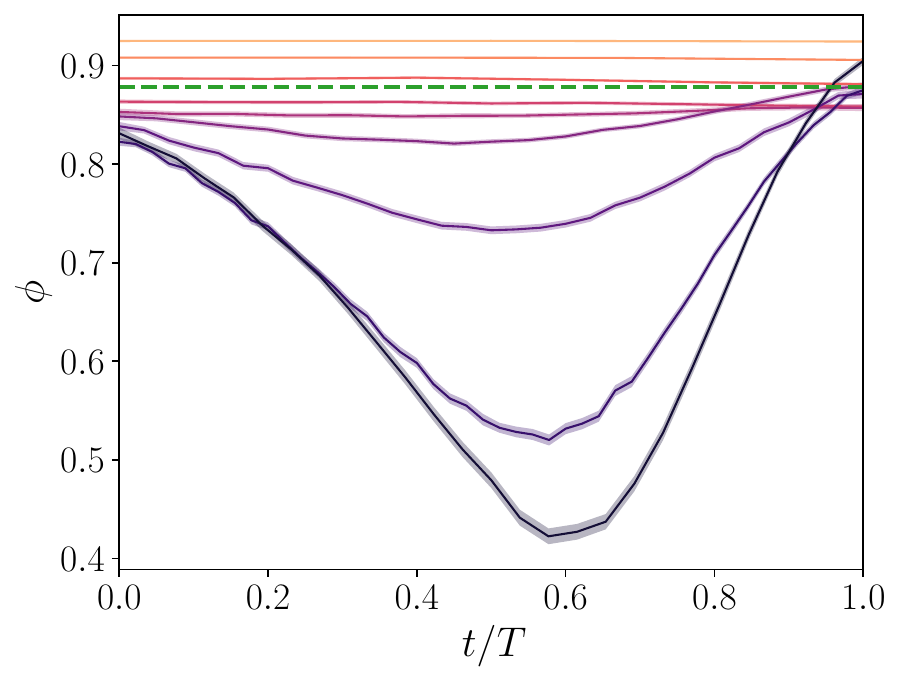}%
}
\subfloat[\label{fig:dynamics:directedDistance}]{%
  \includegraphics[height=0.22\textwidth]{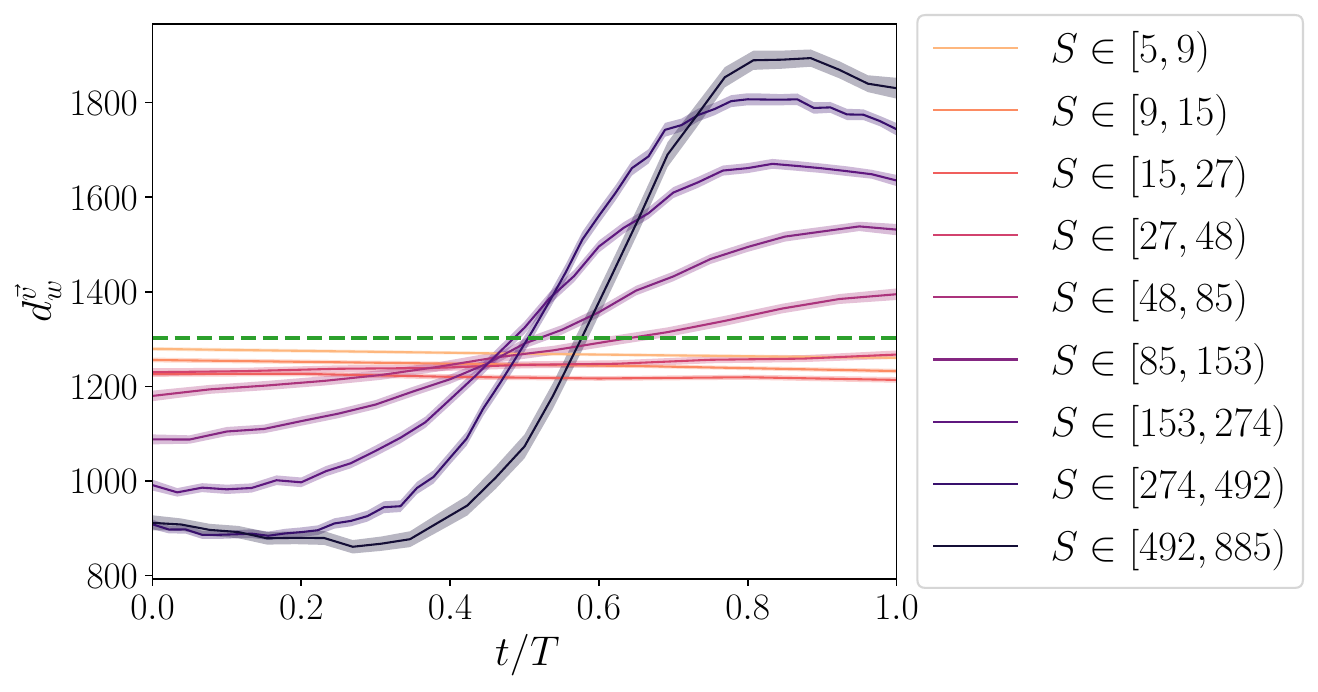}%
}
\caption{Dynamics within turning avalanches of (a) the center of mass speed
  $v_{CM}$, (b) the polarization $\phi$ and (c) the directed wall distance
  $d_w^{\vec{v}}$ depending on the normalized
  time $t/T$ and averaged for similar sizes $S$. The green dashed horizontal line is the average of the given
  variable over the whole experiment.}\label{fig:dynamics}
\end{figure*}

In this section we examine how an avalanche can affect the
behavior of the whole group by measuring several group properties along the evolution of the avalanche.  In order to compare avalanches with different sizes
$S$, as in the case of the avalanche shape discussed above, we first normalize
the temporal evolution of the avalanche by its duration $T$, and then average
the dynamics over groups of avalanches with similar sizes.

First, we investigate the speed of the group given by the center of mass speed,
$v_{CM}$, defined in Eq.~\eqref{eq:3}. We show how it evolves during a turning
avalanche, averaged for different sizes $S$, in Fig.~\ref{fig:dynamics:vCM}. For
comparison, we plot the average value over the whole experiment as the green
dashed horizontal line. We observe that avalanches tend to start below the
average $v_{CM}$, and that avalanches of small size do not alter the school
speed noticeably. On the other hand, larger size avalanches tend to originate at
lower values of $v_{CM}$ and increase the school speed during their evolution.

As a second characteristic of the school we consider the global order measured
in terms of the \textit{polarization} $\phi$~\cite{vicsekNovelTypePhase1995},
\begin{equation*}
  \phi \equiv \left | \frac{1}{N} \sum_i \frac{ \vec{v}_i  }{  v_i } \right|,
  \label{eq:4}
\end{equation*}
which tends to $1$ if the school is ordered and all individuals move in the same
direction, and takes a value close to zero if the school is disordered and fish
move in random and independent directions~\cite{vicsekNovelTypePhase1995}. We
show its evolution within an avalanche in Fig.~\ref{fig:dynamics:phi}. Small
size avalanches tend to start in highly polarized configurations and do not
change significantly the level of order. Contrarily, large avalanches tend to
start with less ordered configurations than the average and further reduce the
order as the avalanche spreads. However, at later stages this trend is reversed
and the school recovers a highly ordered state. 

To gain further information about the possible role of the walls, we
study the dynamical evolution of avalanches with respect to the distance to
the tank walls.  We define the \emph{directed wall distance} $d_w^{\vec{v}}$ as
the distance from the center of mass of the school to the tank walls along the
direction of the velocity of the center of mass.  For a square tank, this
distance is defined as
\begin{align*}
\begin{split}
  d_w^{\vec{v}} \equiv \min \Bigl [ &\sqrt{1+ \left ( \frac{v_{y}}{v_{x}} \right
    )^2 } \left ( \Theta(v_x) (L-x) + \Theta(-v_x) x \right ),\\
  &\sqrt{1+ \left ( \frac{v_{x}}{v_{y}} \right )^2 } \left ( \Theta(v_y) (L-y) +
     \Theta(-v_y) y \right ) \Bigr ],
\end{split}
\end{align*}
where the positions $\vec{x}$ and velocities $\vec{v}$ refer to the center of
mass, $\Theta(x)$ is the Heaviside step function, which discriminates the
forward and backward motion, $L$ is the side of the tank, and the two terms in
the $\min$ function refer to the walls on the $x$ and $y$ coordinates,
respectively. We plot the evolution of this quantity during turning avalanches
in Fig.~\ref{fig:dynamics:directedDistance}. As we can observe, small size
avalanches do not alter the directed wall distance. On the other hand, large
avalanches tend to start closer to the wall and end at higher directed
distances. This indicates that large turning avalanches typically produce a
large change of the group orientation from facing a nearby wall to facing a
farther away wall.  We have also studied the evolution of the distance to the
nearest wall, which we refer as the \emph{minimum wall distance} $d_w$,
\begin{equation*}
  d_w \equiv \min ( x, L-x, y, L-y).
\end{equation*}
We observe (see Supplementary Fig.~\ref{app:fig:dWalls}) that this
quantity decreases and has a minimum for large avalanche sizes, indicating that
during the avalanche evolution the school tends to approach the closest wall, to
later move away from it.

\section{Avalanche correlations}
\label{sec:aval-corr}

Another important aspect in avalanche behavior is the presence of
\textit{correlations}, namely, whether the occurrence of an avalanche induces
the occurrence of other avalanches, such that they appear clustered in space
and/or time~\cite{baroAvalancheCorrelationsMartensitic2014}. The idea of
correlations and clustering in avalanches is closely linked to the concept of
main events and aftershocks in
seismology~\cite{scholzMechanicsEarthquakesFaulting2019}. In this context,
\emph{aftershocks} are typically smaller events that occur after a main event in
nearby locations and stand-out from the background noise. A relevant result here
is the observation of the Omori law, which states that the probability to
observe an aftershock at a given time $t$ after a main event, follows the
distribution
\begin{equation}
  P(t) = \frac{K}{(t+c)^p},\label{eq:omori}
\end{equation}
where $K$, $c$ and $p$ are constants, with
$p \sim 1$~\cite{omoriAftershocksEarthquakes1894}.

In seismology, earthquakes are quantified by their magnitude, which is a measure
related to the logarithm of the energy released. Analogously, for turning
avalanches we can introduce the \emph{magnitude} $m$ as
\begin{equation*}
  m \equiv \ln S,
\end{equation*}
where $S$ is the size of the avalanche. Considering the observed size
distribution from Eq.~\eqref{eq:1}, magnitudes for turning avalanches follow the
distribution
\begin{equation}
  P(m) \sim e^{-b m},
  \label{eq:magnitude}
\end{equation}
with $b = \tau - 1$, which is analogous to the well-known Gutenberg-Richter law
for earthquakes~\cite{gutenbergEarthquakeMagnitudeIntensity1942}.

In order to classify events (either earthquakes or avalanches) into main events
and aftershocks, we consider the method proposed by Baiesi and
Paczuski~\cite{baiesiScalefreeNetworksEarthquakes2004,baiesiComplexNetworksEarthquakes2005}. This
method is based on the definition of the \emph{proximity} $\eta_{ij}$ in
space-time-magnitude domain from an event $j$ to a previous (in time) event
$i$~\cite{baiesiScalefreeNetworksEarthquakes2004,
  zaliapinClusteringAnalysisSeismicity2008,
  zaliapinEarthquakeDeclusteringUsing2020}. Assuming that events are ordered in
time, $t_1 < t_2 < t_3 \cdots$, the proximity is defined as
\begin{equation*}
  \eta_{ij} \equiv
  \begin{cases}
    t_{ij} \,r_{ij}^d \,P(m_i), & \text{if}\ i < j \\
    \infty, & \text{otherwise}
    \end{cases},
  \label{eq:5}
\end{equation*}
where $t_{ij}$ is the time interval between events $i$ and $j$, $r_{ij}$ is the
spatial distance between the events locations, $d$ is the fractal dimension of
the set of events positions and $P(m_i)$ is the Gutenberg-Richter law for event
$i$, which in our case is given by Eq.~\eqref{eq:magnitude}. In the context of
turning avalanches, we have to consider two facts: (i) Avalanches have a finite
duration that is comparable to the inter-event time between consecutive
avalanches. We therefore consider $t_{ij}$, $i < j$, as the number of frames
between the end of avalanche $i$ and the start of avalanche $j$; (ii) During an
avalanche, the school moves. We thus consider the distance $r_{ij}$, $i < j$, as
the distance between the center of mass of the school at the end of avalanche
$i$ and the center of mass of the school at the beginning of avalanche
$j$. Additionally, the distribution of the positions of the center of mass at
the start of avalanches does not seem to show a fractal structure, so we use
here $d=2$.

The proximity $\eta_{ij}$ is a measure of the expected number of events  of magnitude $m_i$ to occur,
 looking backward in time from event $j$ within a time
interval $t_{ij}$ and distance $r_{ij}$, in the absence of correlations, in such a way
that the time and position of previous avalanches behave as independent Poisson
processes~\cite{baiesiScalefreeNetworksEarthquakes2004}.   
Therefore, smaller values of the proximity are associated to a larger probability that the events $i$ and $j$ are actually correlated.

\begin{figure}[t!p]
\centering
\subfloat[\label{fig:aftershocks:nAftershocks}]{%
  \includegraphics[width=0.23\textwidth]{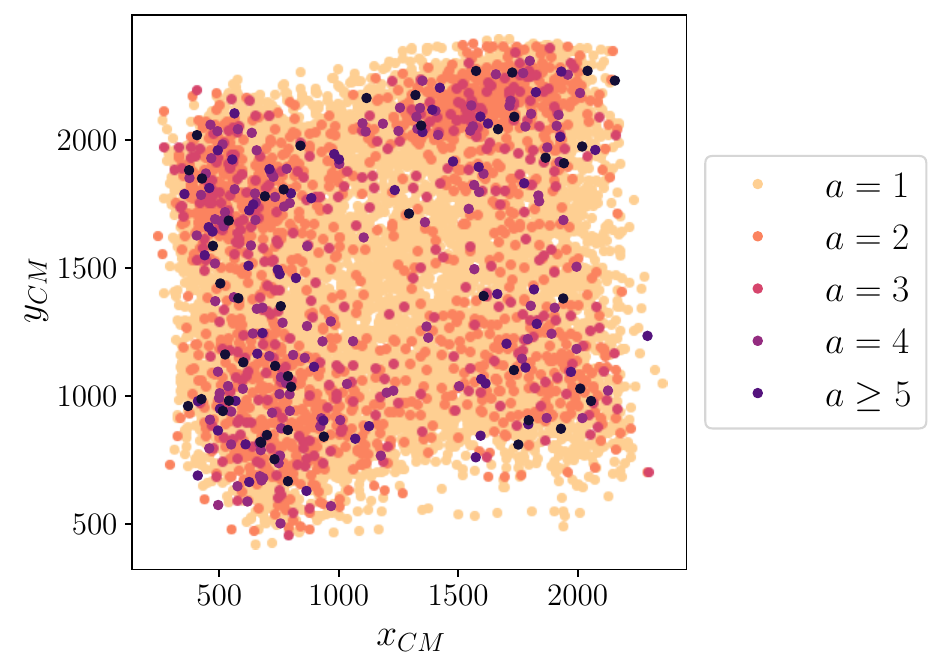}%
}
\subfloat[\label{fig:aftershocks:rij_tij}]{%
  \includegraphics[width=0.22\textwidth]{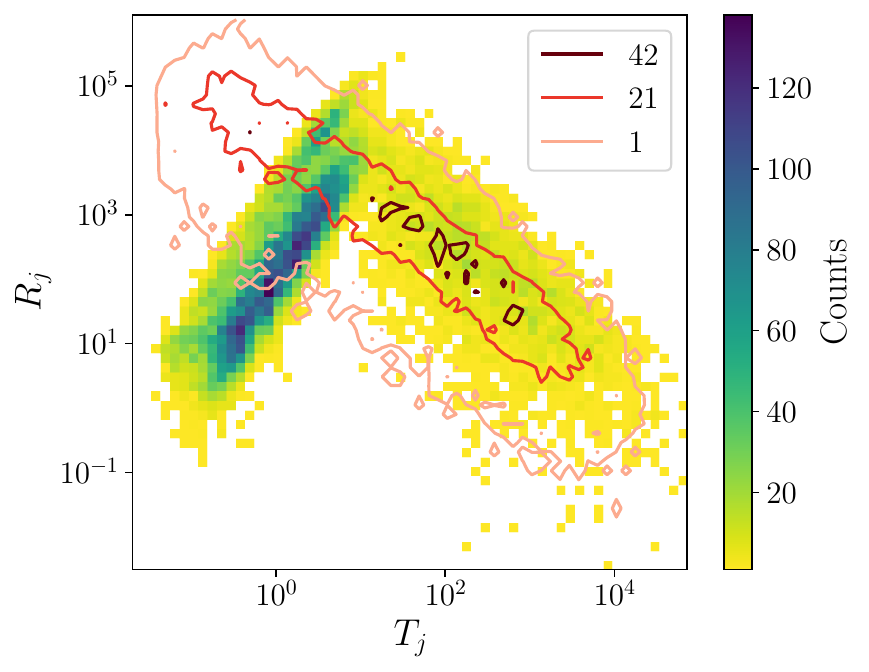}
}
\hspace{0.01\textwidth}
\subfloat[\label{fig:aftershocks:P_tij}]{%
\includegraphics[width=0.22\textwidth]{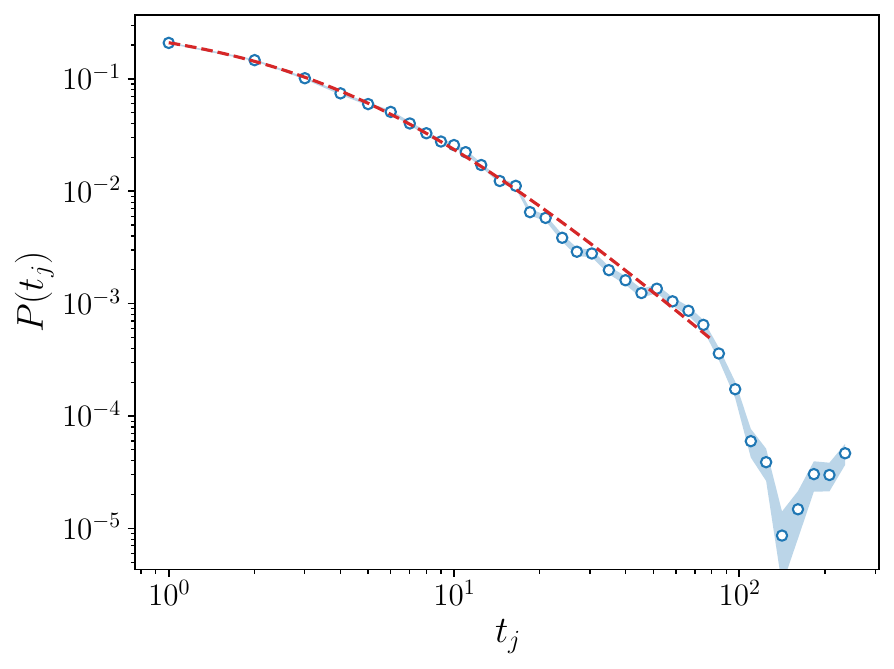}
}

\caption{Correlation measures of aftershocks. (a) number of aftershocks $a$ per parent
  depending on the triggering location of the parent, (b) counts for the joint
  distribution of the rescaled space $R_{j}$ and time $T_{j}$ (the contour plot
  corresponds to randomized avalanches, in which avalanche positions, inter-event times and magnitudes have been
  shuffled) and (c) PDF for the time interval $t_{j}$ between parents and
  aftershocks for $t_{j} < 250$. We only considered
  avalanches with magnitudes $m\ge 1.6$. In (c), the red dashed line
  corresponds to a fit to the Omori law Eq.~\eqref{eq:omori} with $c=4.3\pm0.4$ and
  $p=2.2\pm0.1$.}\label{fig:aftershocks}
\end{figure}

Using the  proximity $\eta_{ij}$, every event $j$
can be associated to a \textit{nearest neighbour} or \emph{parent} $p_j$,
defined as the event in the past ($p_j < j$) that
minimizes the proximity with $j$, namely $\eta_{p_jj} \leq \eta_{ij}$, $\forall i < j$.  This proximity is denoted the
\textit{nearest-neighbour proximity} $\eta_j$, its time interval $t_j$ and the
spatial distance $r_j$. The set of events with the same parent are considered
the aftershocks of that parent. In Fig.~\ref{fig:aftershocks:nAftershocks} we
examine the distribution of the triggering locations of parents, color-coded by their number of aftershocks $a$. We find a possible influence of the tank walls,
as parents with larger number of aftershocks tend to be located nearer the
corners.

In addition, we consider the measure of clustering proposed within this
framework in Ref.~\cite{zaliapinClusteringAnalysisSeismicity2008}. This
formalism is based in the \emph{rescaled time} $T_{j}$ and \emph{rescaled
  space} $R_{j}$~\cite{zaliapinClusteringAnalysisSeismicity2008,
  zaliapinEarthquakeDeclusteringUsing2020}, defined as
\begin{align*}
  T_{j} &\equiv t_{j}\sqrt{P(m_{p_j})},\\
  R_{j} &\equiv (r_{j})^d\sqrt{P(m_{p_j})},
\end{align*}
such that
\begin{equation*}
  \eta_{j} =T_{j} R_{j}.
\end{equation*}
In real earthquakes, it is observed that the joint distribution of $T_{j}$ and
$R_{j}$ is bimodal. One mode corresponds to background events, and is compatible
with a random (Poisson) distribution of times and positions of events. The other
mode, on the other hand, corresponds to clustered events, correlated in space
and time~\cite{zaliapinEarthquakeDeclusteringUsing2020}.

In Fig.~\ref{fig:aftershocks:rij_tij} we show the joint distribution of $T_{j}$
and $R_{j}$ for turning avalanches in terms of a color density plot. In the same
figure, we display in terms of a contour plot the joint distribution obtained
for randomized data, in which avalanche positions, inter-event times and
magnitudes have been shuffled. We find that the experimental data shows clearly
 two modes in the distribution. In one mode, for large values of $T_{j}$,
increasing the rescaled time $T_{j}$ results in a decrease of the rescaled space
$R_{j}$. This is almost identical to the distribution obtained for the shuffled
data, indicating that it corresponds essentially to background, uncorrelated noise. The other
mode occurs for smaller values of $T_{j}$ and displays the opposite behaviour,
increasing the rescaled time $T_{j}$ results in a higher rescaled space
$R_{j}$. This behaviour is different from the background noise and corresponds
to clustered (correlated) avalanches.

We can understand the time scale separation between the modes taking into
account that turning avalanches take place inside a school that is moving around
the tank. The school typically performs a recurrent movement on the tank,
visiting a given point in the tank with some average period. We can
quantitatively analyse this behaviour looking at the mean square displacement of the
position of the center of mass, which measures the average displaced distance of the group in time starting from any point in the trajectory (see Supplementary
Fig.~\ref{app:fig:autocorr_xCM}). The first maximum occurs around $t_c = 250$
frames and corresponds to the average time the school needs to perform a
half-turn around the tank and becomes maximally separated from its initial
position. Aftershocks with a lower time interval tend to increase their spatial distance as the school moves away from the parent location. After this time and up to very large time intervals, the school may return towards the
parent position and we can find aftershocks occurring at lower spatial
distances. However, these tend to occur rather randomly and can not be
distinguished from random events. This highlights a major difference with
earthquakes, where significant correlations can occur in the same location at
widely separated time intervals.

Finally, we examine the Omori law displaying the distribution for the time
interval $t_j$ between parents and aftershocks in
Fig.~\ref{fig:aftershocks:P_tij}. The distribution is computed considering the sequences of aftershocks for each parent, shifting the sequences to  set each parent at a common time zero, and stacking all sequences in a single common sequence~\cite{ouillonMagnitudedependentOmoriLaw2005}.
From the above reasons, we only consider time
intervals below $t_c = 250$ that correspond to significant correlated
aftershocks.
A least-squares fitting of the empirical data to the Omori law given by
Eq.~\eqref{eq:omori} (green dashed line), yields the parameters $c = 4.3 \pm 0.4$
and $p = 2.2 \pm 0.1$. This indicates a value $p > 1$, implying a faster decay
rate of aftershocks than in earthquakes.

\section{Discussion}

In this paper, we have presented an empirical analysis of spontaneous behavioral
cascades in schooling fish considering turning avalanches, where large turns in
the direction of motion of individuals are propagated across the
group. This was achieved collecting extensive, state-of-the-art
  tracking data for schooling fish, comprising up to $1.8 \times 10^5$ time samplings
  at a resolution of 50 frames per second, for experiments involving varying
  numbers of fish, up to groups of 50 individuals. This dataset yielded over
  $10^4$ avalanche events, representing a significant advancement compared to
  previous studies on behavioral cascades (for reference,
  in~\cite{poelSubcriticalEscapeWaves2022} the authors reported $10^2$
  avalanche events). We have analyzed different avalanche metrics and
provided a highly detailed picture of the dynamics associated to
  behavioural cascades, employing tools from avalanche behavior in condensed
matter physics and seismology.

We have uncovered  evidence of scale-free behavior across various aspects of turning avalanches in schooling fish. Analysis of probability distributions for fundamental observables, such as the avalanche duration, size, and inter-event times, revealed long tails compatible with power-law forms. Adjusting for dragon king events, which are disproportionately represented extreme events induced by interactions with tank walls, we found the power-law region for avalanche size extended up to two decades. We also established a scaling relationship between the characteristic exponents of the duration and size distributions. Furthermore, a data collapse in the distributions of the duration and inter-event times at a fixed activity rate, indicates a connection in avalanche dynamics across schools with varying number of individuals and the turning threshold defining the avalanche. We also confirmed two previously observed data collapses in critical avalanche systems: in the inter-event times distribution normalized by the mean, and in the avalanche shape or mean temporal profile via a scaling relationship with the duration.



While power laws are often attributed to critical phenomena related to
  phase transitions, alternative mechanisms can also produce such
  distributions~\cite{sornetteCriticalPhenomenaNatural2004,
    munozColloquiumCriticalityDynamical2018}. A tighter prediction of
  criticality is manifested through data collapses and relations between scaling
  exponents~\cite{munozColloquiumCriticalityDynamical2018,
    friedmanUniversalCriticalDynamics2012, sethnaCracklingNoise2001}, indicating
  quantitative universal avalanche dynamics across scales. Ultimately, these
  findings are insufficient to demonstrate criticality, but they constitute
  necessary conditions and embody a crucial theoretical aspect that has received
  limited attention in behavioral cascades of moving animal groups. In our work,
  we address this gap and complement existing studies, providing evidence that
  fish schools could operate in the vicinity of a critical point. In particular,
  given the apparent lack of externally tuned parameters in the system, they
  would represent an instance of self-organized
  criticality~\cite{romanczukPhaseTransitionsCriticality2022,
    tadicSelfOrganisedCriticalDynamics2021}. While our experimental settings are currently limited to studying small group sizes of up to 50 individuals, future work should aim to test for criticality effects in larger groups, involving hundreds or thousands of individuals, to draw adequate comparisons with statistical physics systems.

Being near a critical point can offer advantages such as efficient collective decision-making and information transfer across the group~\cite{moraAreBiologicalSystems2011, munozColloquiumCriticalityDynamical2018, romanczukPhaseTransitionsCriticality2022, gomez-navaFishShoalsResemble2023}. In this context, turning avalanches may arise from self-organized critical processes that facilitate information exchange among members of a social system, comparable to avalanches seen in the social interactions of collective knowledge creation~\cite{dankulovDynamicsMeaningfulSocial2015, tadicMechanismsSelforganizedCriticality2017}. Specifically, turning avalanches allow fish to decide collectively on their direction of movement. For this reason, it is not surprising that we observe large avalanches occurring at the onset of the active phase of the burst-and-coast mechanism in fish locomotion, where decision-making processes to change individual directions are believed to occur~\cite{herbert-readHowPredationShapes2017, harpazDiscreteModesSocial2017, caloviDisentanglingModelingInteractions2018}. During the process of deciding a new collective direction, coordination and group order decrease. However, once a new direction is chosen, speed increases, and coordination re-emerges. A similar behavior was observed in the phenomenon of collective U-turns, involving directional switches for fish swimming in a ring-shaped tank~\cite{jiangIdentifyingInfluentialNeighbors2017, lechevalSocialConformityPropagation2018}. We argue that collective U-turns can be understood as a specific example of turning avalanches.


Boundary effects, arising from interactions with tank walls or distinct
behaviors of individuals at the group's border, are frequently overlooked in the
study of animal collective motion. This work highlights significant effects of
tank walls on avalanche behavior. While walls do not increase the number of
avalanches, those in their proximity often exhibit larger sizes and manifest in
correlated clusters, resulting in a higher occurrence of
aftershocks. Moreover, individuals that are initiators of avalanches
  are more frequently found near walls. This phenomenon can be attributed to the
  tank walls acting as obstacles, disrupting the group's movement and prompting
collective decisions for a subsequent direction away from the
walls~\cite{joshiAlignmentNeighboursEnables2022, wuExperimentalStudyDecisionmaking2024}. Notably, large avalanches
induced by tank walls primarily impact the tail of duration, size, and
inter-event time distributions, manifesting as shoulders or dragon
  kings. The intermediate scale-free behavior in these distributions appears to
  be intrinsic to spontaneous turning avalanche mechanisms, rather than being promoted
  by the walls. Additionally, boundary effects from individuals at the group's
border play a role, as they are often initiators of avalanches. This aligns with
previous findings associating these positions with higher social
influence~\cite{cavagnaBoundaryInformationInflow2013,
  rosenthalRevealingHiddenNetworks2015}. An alternative explanation is
  that individuals at the group's border may be more exposed to
risks~\cite{hamiltonGeometrySelfishHerd1971}, maintaining a heightened alert
state and making them more prone to initiating a large change of direction.


We have examined the spatial and temporal correlations in turning
avalanches through the concept of aftershocks~\cite{baroAvalancheCorrelationsMartensitic2014}. 
We observe that  turning
avalanches of schooling fish reveal significant clustered and correlated events
 below a time interval corresponding to a half-turn of the school around the
tank. This observation points to a fundamental property linked to the absence of
collective memory for larger time
scales~\cite{couzinCollectiveMemorySpatial2002}. Furthermore, we found that the
probability rate of observing correlated aftershocks after a main event in
turning avalanches follows an Omori law with a decay rate exponent $p \sim 2$,
 significantly faster than in seismology ($p \sim 1$).


We believe this work makes a contribution to the ongoing inquiry into criticality, particularly within the realm of animal collective motion and, more broadly, in biological systems. The limited number of analyses conducted on large datasets with experimental evidence of self-similar behavior—a hallmark of critical systems—highlights the need for further exploration and clarification in this area. Future experiments should aim to study larger systems over longer periods of time to deepen our understanding of these phenomena.



\begin{acknowledgments}
We acknowledge financial support from projects PID2022-137505NB-C21 and PID2022-137505NB-C22 funded by MICIU/AEI/10.13039/501100011033, and by ``ERDF A way of making Europe''. A. P. acknowledges a fellowship from the Secretaria
  d’Universitats i Recerca of the Departament d’Empresa i Coneixement,
  Generalitat de Catalunya, Catalonia, Spain.
We thank P. Romanczuk, H. J. Herrmann, and E. Vives for helpful
  comments.
\end{acknowledgments}

{
\appendix

\section{Experimental data}
\label{sec:methods:experimentalData}

We employ schooling fish of the species black neon tetra (\textit{Hyphessobrycon
  herbertaxelrodi}), a small freshwater fish of average body length 2.5~cm that
has a strong tendency to form cohesive, highly polarized and planar
schools~\cite{gimenoDifferencesShoalingBehavior2016}. The experiments,
performed at the Scientific and Technological Centers UB (CCiTUB), University of
Barcelona (Spain), were reviewed and approved by the Ethics
Committee of the University of
Barcelona (project number 119/18). They involved schools of $N=8, 16, 32$ and $50$ individuals
freely swimming in a square tank of side $L=100$ cm with a water column of
$5$~cm of depth, resulting in an approximately two-dimensional movement. Videos
of the fish movement were recorded with a digital camera at 50 frames per
second, with a resolution of $5312\times2988$ pixels per frame, the side of the
tank measuring $L=2730$ pixels. Digitized individual trajectories were obtained
from the video recordings using the open source software
idtracker.ai~\cite{romero-ferreroIdtrackerAiTracking2019}.  Invalid values
returned by the program caused by occlusions were corrected in a supervised way,
semi-automatically interpolating with spline functions (now incorporated in the
Validator tool from version 5 of idtracker.ai). For better accuracy, we
projected the trajectories in the plane of the fish movement, warping the tank
walls of the image into a proper square (for details
see Ref.~\cite{puySelectiveSocialInteractions2024}). We smoothed the trajectories
with a Gaussian filter~\cite{nixonFeatureExtractionImage2010} with $\sigma = 2$ and truncating the filter at $5\sigma$, employing the \texttt{scipy.ndimage.gaussian\_filter1d function  from the 
\texttt{scipy} Python scientific library~\cite{virtanenSciPyFundamentalAlgorithms2020}}. Individual velocities and accelerations were obtained from the Gaussian
filter using first and second derivatives of the Gaussian kernel, respectively.

We discarded recordings where fish stop for prolonged periods. We implement this quantitatively applying a Gaussian filter with $\sigma = 200$ to the mean speed of individuals $\left< v \right >$ and discarding sequences that go below a given threshold $\left< v \right >_{th} = 1.5$. The remaining experiments we analyze consist in 6 independent recordings (performed on
different days and with different individuals) of $N=8$ fish during $30$~min
(90000 frames), 3 recordings of $N=16$ fish during $30$~min, 3 recordings of $N=32$ fish during $30$~min and 3 recordings of
$N=50$ fish during $60$~min (180000 frames).  The data with $N=8$ fish was
previously used in Ref.~\cite{puySelectiveSocialInteractions2024}.



\section{Turning rate formula}
\label{sec:methods:turning rate formula}

The turning rate $\omega$ is defined as the absolute value of the rate of change of the orientation $\theta$ of the velocity of an individual with time, i.e.
\begin{equation*}
  \omega \equiv \left |\frac{d\theta}{dt} \right|.
\end{equation*}
Consider the velocity vector in two instants of time, $t$ and $t + \Delta
t$. The change of orientation $\Delta \theta$ from $\vec{v}(t)$ to $\vec{v}(t+\Delta t)$ is given by
\begin{equation*}
  \label{eq:11}
  \sin (\Delta \theta ) = \frac{\vec{v}(t) \times \vec{v}(t+\Delta t)
    }{v(t+\Delta t) v(t)}.
\end{equation*}
In the limit $\Delta t \to 0$, $\Delta \theta \to 0$, we have
\begin{eqnarray*}
  \label{eq:12}
  \sin (\Delta \theta ) &\simeq& \frac{1}{v(t+\Delta t) v(t)} \left \{
    \vec{v}(t) \times \left [\vec{a}(t) \Delta t + \vec{v}(t) \right] \right \} =
                         \nonumber \\
  &=& \frac{ \vec{v}(t) \times \vec{a}(t) 
    }{v(t+\Delta t) v(t)} \Delta t \simeq \Delta \theta,
\end{eqnarray*}
where $\vec{a}(t)$ is the fish acceleration. Then we can write
\begin{equation*}
  \frac{d\theta}{dt} =  \lim_{\Delta t \to 0} \frac{\Delta \theta}{\Delta t}  = \lim_{\Delta t
    \to 0}  \frac{ \vec{v}(t) \times \vec{a}(t) 
    }{v(t+\Delta t) v(t)} 
    = \frac{\vec{v}(t) \times  \vec{a}(t) 
    }{v(t)^2},\label{eq:13}
\end{equation*}
recovering the expression for the turning rate in Eq.~\eqref{eq:6}.

\section{Null model of avalanches without temporal correlations} 
\label{sec:methods:null_model}

Following Ref.~\cite{mugicaScalefreeBehavioralCascades2022}, we can consider a
null model of avalanche behavior in schooling fish in which individuals perform
random uncorrelated turning rates, extracted from the empirical distribution
$P(\omega)$. In this case, the probability $q$  that, at a given frame, a fish performs a turning
rate larger than a threshold $\omega_{th}$ (i.e. a fish is active) is given by
\begin{equation*}
  q = \int_{\omega_{th}}^{\infty} P(\omega) \; d\omega,
  \label{eq:sup_turningprobn}
\end{equation*}
while the probability that, at a given frame, at least one fish in a school of
$N$ individuals performs a turning rate larger than $\omega_{th}$ (i.e. there
is at least one active fish) is
\begin{equation*}
  Q = 1 - (1-q)^N.
\end{equation*}
In this null model, an avalanche of duration $T$ implies $T$ consecutive frames
with at least an active fish, followed by a frame with no active fish. Thus the
duration distribution has the normalized form
\begin{equation}
  \label{eq:7}
  P_0(T) =  \frac{1-Q}{Q} Q^{T},\;\;\; T \in [1, \infty).
\end{equation}
An inter-event time $t_i$ consists, analogously, of $t_i$ consecutive frames
with no active fish, followed by a frame with at least one active
fish. Therefore the inter-event time distribution has the form
\begin{equation}
  \label{eq:17}
  P_0(t_i) =  \frac{Q}{1-Q} (1-Q)^{t_i},\;\;\; t_i \in [1, \infty).
\end{equation}
Finally, the size distribution can be estimated as
follows~\cite{mugicaScalefreeBehavioralCascades2022}: At each frame during an
avalanche, the average number of active fish is $N q / Q$, where the
normalization factor $Q$ accounts for the fact that at least one fish was active
in the frame considered. Thus, an avalanche of duration $T$ has an average size
$S = TNq/Q$. Transforming the duration distribution Eq.~\eqref{eq:7}, we then
have~\cite{mugicaScalefreeBehavioralCascades2022}
\begin{equation*}
  \label{eq:8}
  P_0(S) = \frac{1-Q}{Nq} Q^{\frac{QS}{Nq}}.
\end{equation*}
In all cases, we recover distributions with an exponentially decaying form.


Now, the activity rate $r$, defined as the probability that a randomly chosen
frame belongs to an avalanche, is equal to the probability that in a randomly
chosen frame there is at least one active fish. This trivially implies
\begin{equation*}
  \label{eq:10}
  r = Q.
\end{equation*}
That is, the duration and inter-event time distributions depend only on the
activity rate, and can be made to collapse for different values of $N$ and
$\omega_{th}$ leading to the same value of $r$. On the other hand, the size
distribution depends additionally on $N$ and $q$ and thus cannot be made to
collapse by fixing $r$.

For the inter-event time distribution Eq.~\eqref{eq:17}, we can write, in the
limit of small $Q$,
\begin{equation*}
  \label{eq:15}
  P_0(t_i) \simeq  Q (1-Q)^{t_i} = Q e^{t_i \ln(1-Q)} \simeq Q e^{-Q t_i}.
\end{equation*}
From Eq.~\eqref{eq:17}, $\av{t_i} = \sum_{t_i = 1}^\infty t_i P_0(t_i) = 1/Q$. Thus,
we have
\begin{equation*}
  \label{eq:14}
   P_0(t_i) \simeq \frac{1}{\av{t_i}} e^{-t_i / \av{t_i}},
\end{equation*}
recovering the scaling relation Eq.~\eqref{eq:dataCollapse_interevent} with
$\Phi(x) = e^{-x}$, in the limit of large $\av{t_i}$.

Interestingly, the activity rate $r$ in this null model follows the empirical
behavior shown in Fig.~\ref{fig:rate:avalancheRate}, as $Q$ is a growing
function of $N$ and a decreasing function of $\omega_{th}$.

\section{Statistical dragon kings detection test}\label{sec:methods:statistical_dragon_kings_test}

The statistical dragon kings detection test developed in~\cite{janczuraBlackSwansDragonkings2012}, and also employed in~\cite{mikaberidzeDragonKingsSelforganized2023}, uses a $p$-value to quantify the presence of a dragon king peak in the tail of a heavy-tailed distributed variable $x$. First, the complementary cumulative distribution function  $1-F(x) \equiv P(X \geq x)$ is constructed and the most overrepresented data point in the tail, denoting potential dragon king events, is identified. Next, a power law $ax^b$ is fitted to the appropriate scale-free region of $1-F(x)$. Confidence intervals $\gamma$ of the power law fit are calculated from~\cite{janczuraBlackSwansDragonkings2012}:
\begin{equation*}
\left( \frac{1}{N} q_{1-\gamma/2}(N, a x^b), \frac{1}{N} q_{\gamma/2}(N, a x^b) \right],
\end{equation*}
where $q_{\alpha}(n, z)$ is the $\alpha$-quantile of the binomial distribution $B(n,z)$ and $N$ is the number of elements $x$. Finally, the tightest confidence interval $\gamma^*$ of the power law fit accommodating for the most overrepresented data point in the tail is obtained by visual inspection. The $p$-value of the null hypothesis that there are no dragon kings in the distribution corresponds to $p \equiv 1-\gamma^*$.
}

\bibliography{all}


\clearpage

\newpage

\renewcommand{\thepage}{\arabic{page}} 
\renewcommand{\theequation}{S\arabic{equation}} 
\renewcommand{\thesection}{\arabic{section}}  
\renewcommand{\thetable}{S\arabic{table}}  
\renewcommand{\thefigure}{S\arabic{figure}}
\renewcommand{\thevideo}{S\arabic{video}}


\setcounter{page}{1}
\setcounter{section}{0}
\setcounter{figure}{0}
\setcounter{equation}{0}

\onecolumngrid
\begin{center}
\textbf{\large Supplemental Material\\~\\}
\end{center}

\section{Supplementary Videos}
\label{sec:supplementary-videos}

\begin{video*}[h]
  \caption{Examples of large turning avalanches in schooling fish of $N=50$
    individuals. The trackings are overlapped to the experimental video. In grey
    we display individuals that have not participated yet in the avalanche, in
    cyan the individuals that are active at the current frame, and in blue
    individuals that were active previously in the
    avalanche.\label{app:video:1}}
\end{video*}


\clearpage

\section{Supplementary Figures}
\label{sec:suppl-figur}

\begin{figure*}[h]
\subfloat[\label{app:fig:measures:duration}]{%
  \includegraphics[width=0.4\textwidth]{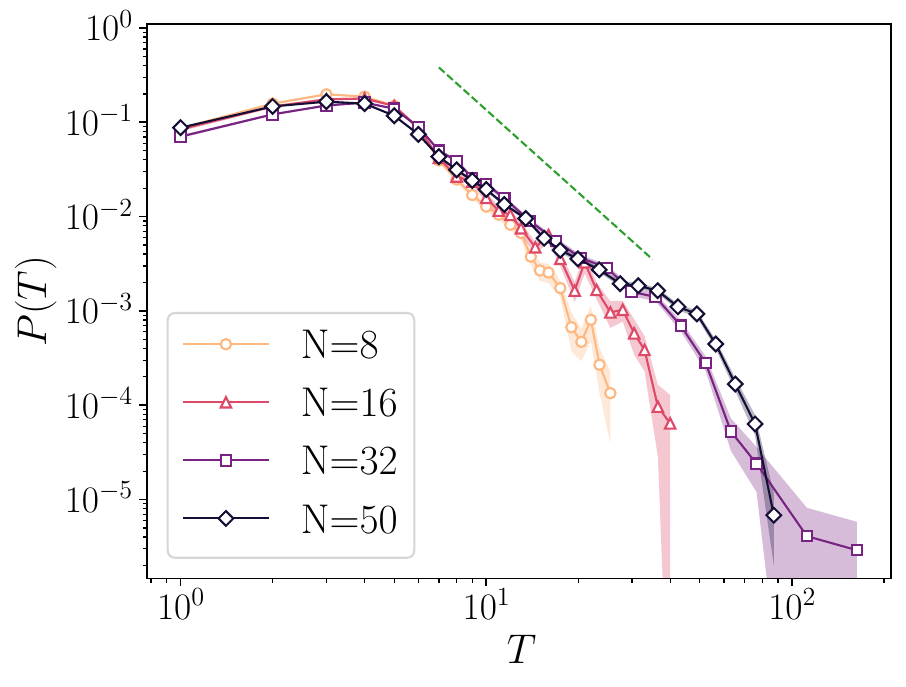}%
}
\hspace{0.01\textwidth}
\subfloat[\label{app:fig:measures:size}]{%
  \includegraphics[width=0.4\textwidth]{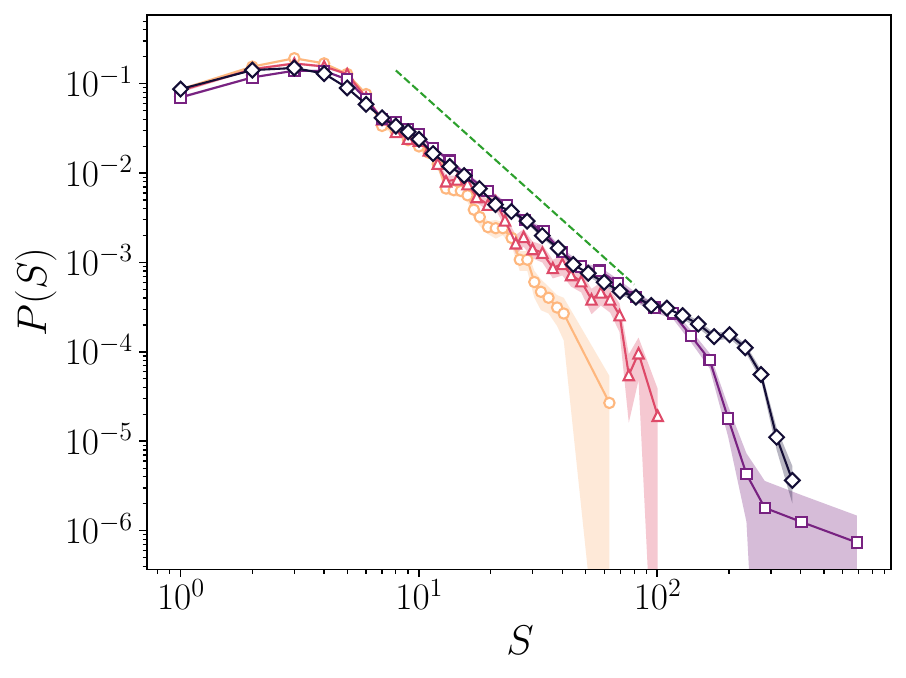}%
}

\subfloat[\label{app:fig:measures:T-S}]{%
  \includegraphics[width=0.4\textwidth]{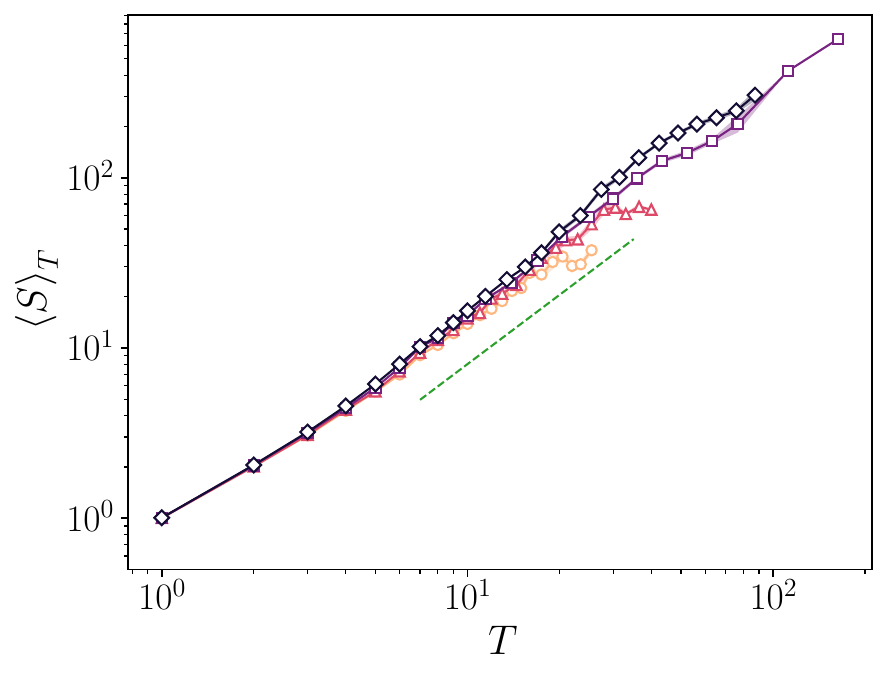}%
}
\hspace{0.01\textwidth}
\subfloat[\label{app:fig:measures:interEvent}]{%
  \includegraphics[width=0.4\textwidth]{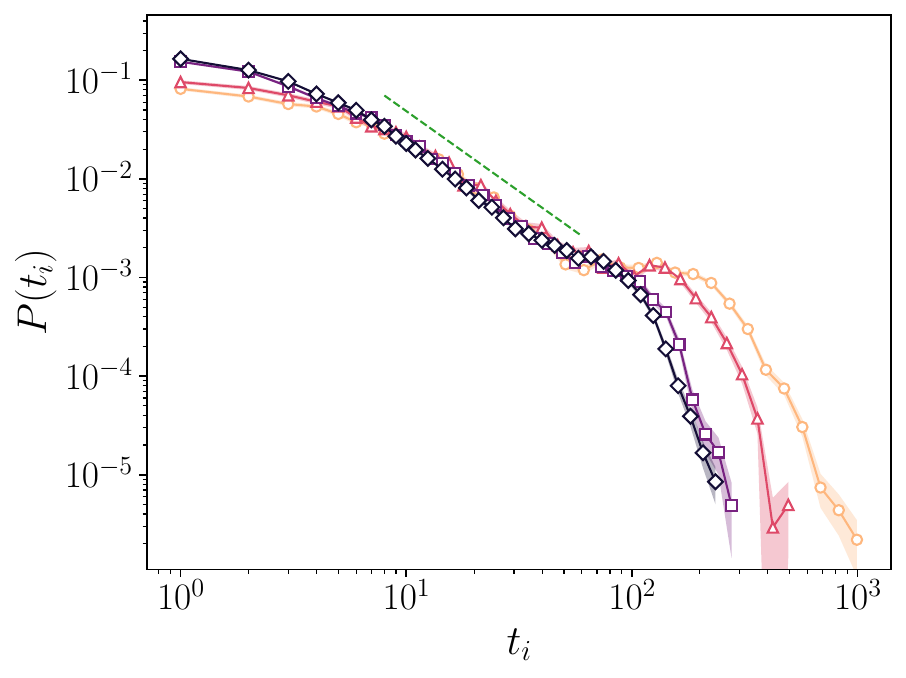}%
}
\caption{(a) PDF of the duration $T$, (b) PDF of the size $S$, (c) average size $\left < S\right >_T$ depending on the duration $T$ and (d) PDF of the inter-event time $t_i$ for $\omega_{th}=0.15$. The different curves correspond to schools of different number of individuals $N$. The exponents from the green dashed power laws are (a) $\alpha = 2.9 \pm 0.8$, (b) $\tau=2.4 \pm 0.4$, (c) $m=1.35\pm 0.16$ and (d) $\gamma = 1.63\pm 0.04$.}\label{app:fig:measures}
\end{figure*}

\begin{figure*}[t!p]
\subfloat[\label{app:fig:centralHist:duration}]{%
  \includegraphics[width=0.4\textwidth]{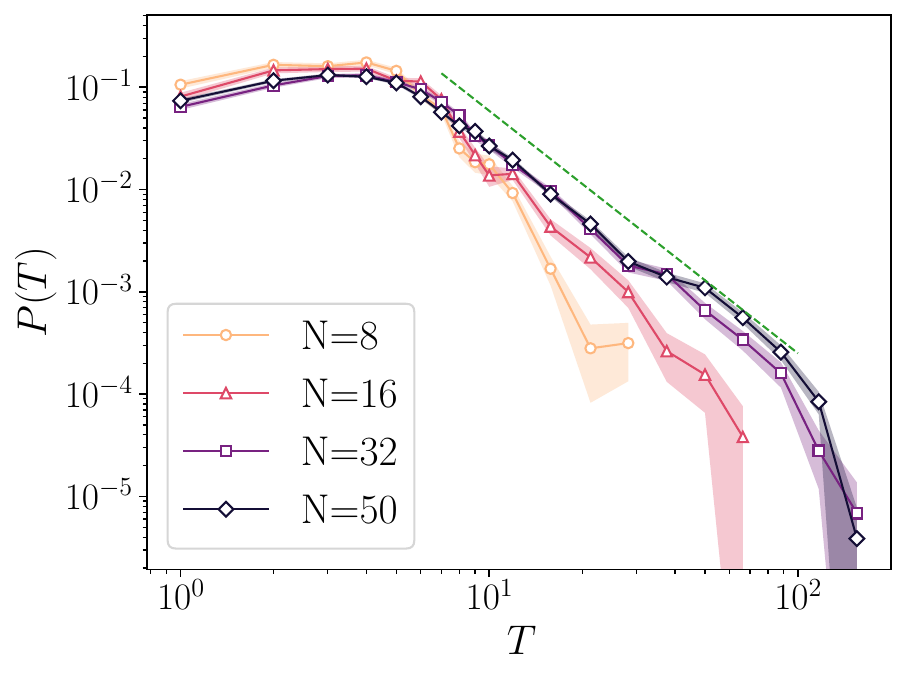}%
}
\hspace{0.01\textwidth}
\subfloat[\label{app:fig:centralHist:size}]{%
  \includegraphics[width=0.4\textwidth]{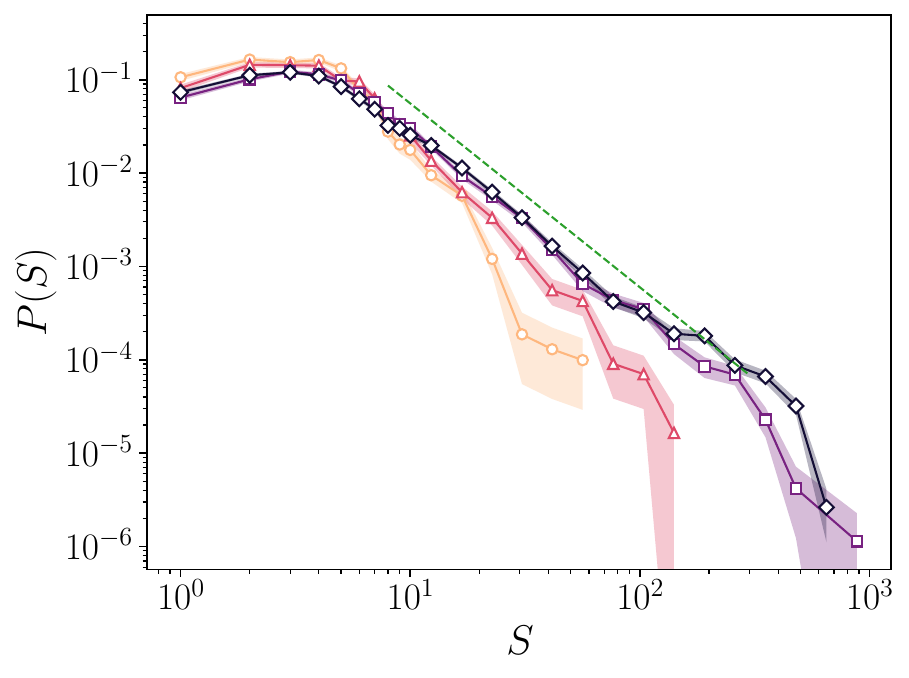}%
}

\subfloat[\label{app:fig:centralHist:T-S}]{%
  \includegraphics[width=0.4\textwidth]{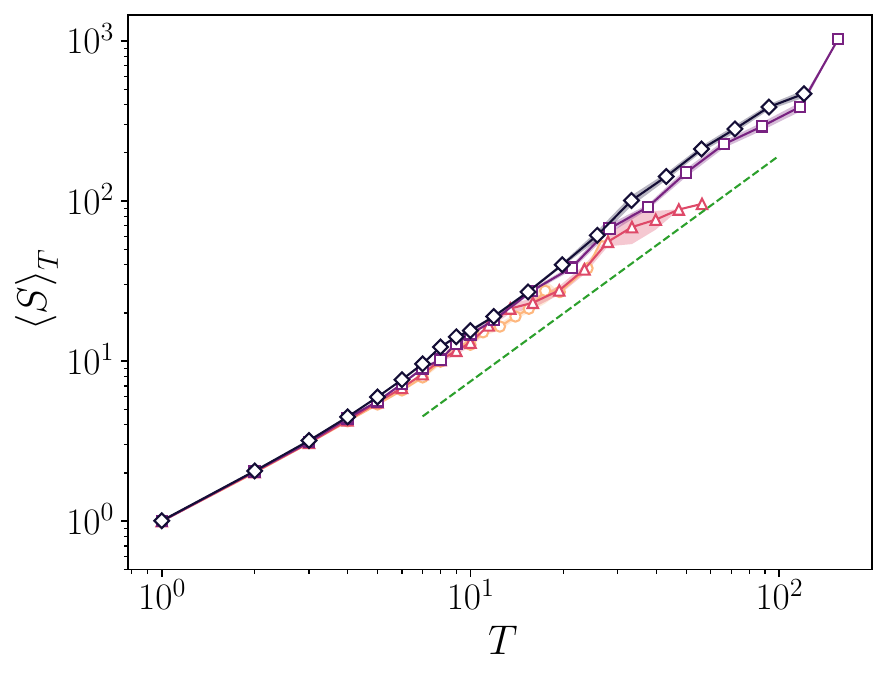}%
}
\caption{(a) PDF of the duration $T$, (b) PDF of the size $S$ and (c) average size $\left < S\right >_T$ depending on the duration $T$ for $\omega_{th}=0.1$ and for triggering locations of avalanches in the central square of side $L/3$, with $L$ the side of the tank. The different curves correspond to schools with different number of individuals $N$. The exponents from the green dashed power laws are the same as in Fig.~\ref{fig:measures}.}\label{app:fig:centralHist}
\end{figure*}

\begin{figure*}[t!p]
  \includegraphics[width=0.43\textwidth]{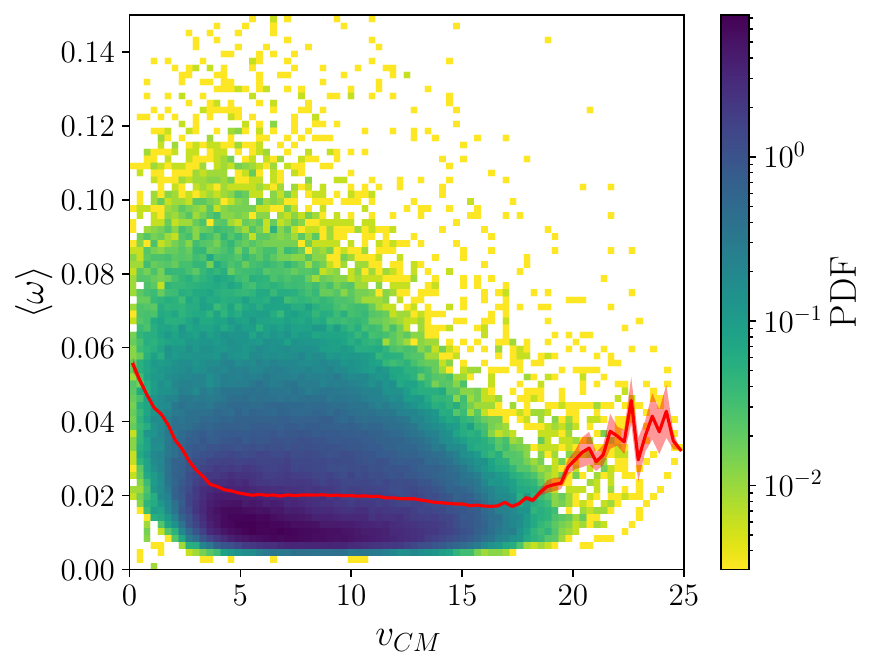}%
  \caption{PDF of the turning rate $\left< \omega \right>$ averaged for all
    individuals depending on the center of mass speed $v_{CM}$. The red line is
    the average of $\left< \omega \right>$ for each value of the speed
    $v_{CM}$.}\label{app:fig:omega-v}
\end{figure*}

\begin{figure*}[t!p]
  \includegraphics[width=0.43\textwidth]{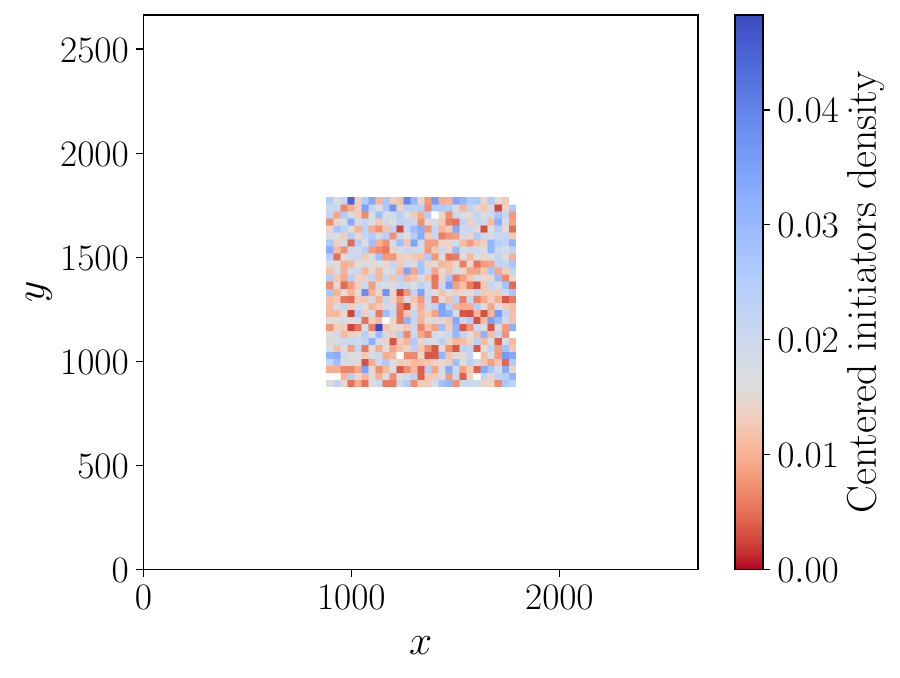}%
\caption{Density for the position of centered initiators normalized against the positions of all centered individuals at the start $t_0$ of an avalanche in the laboratory reference frame. The grey colour in the colormap corresponds to
  the expected density in the absence of correlations.}\label{app:fig:lab_centred}
\end{figure*}

\begin{figure*}[t!p]
  \includegraphics[width=0.6\textwidth]{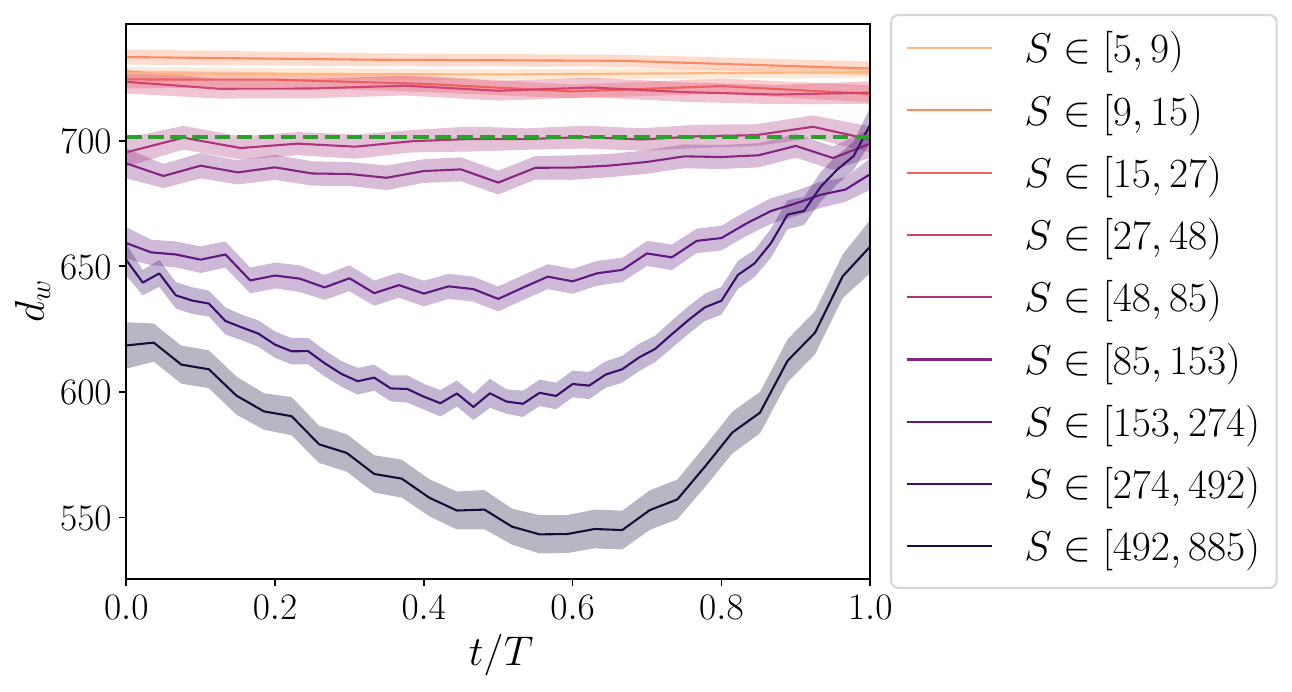}%
\caption{Dynamics within turning avalanches of the minimum wall distance $d_w$ depending on the normalized time $t/T$ and averaged for similar sizes $S$. The green dashed line is the average over the whole experiment.}\label{app:fig:dWalls}
\end{figure*}

\begin{figure*}[t!p]
  \includegraphics[width=0.4\textwidth]{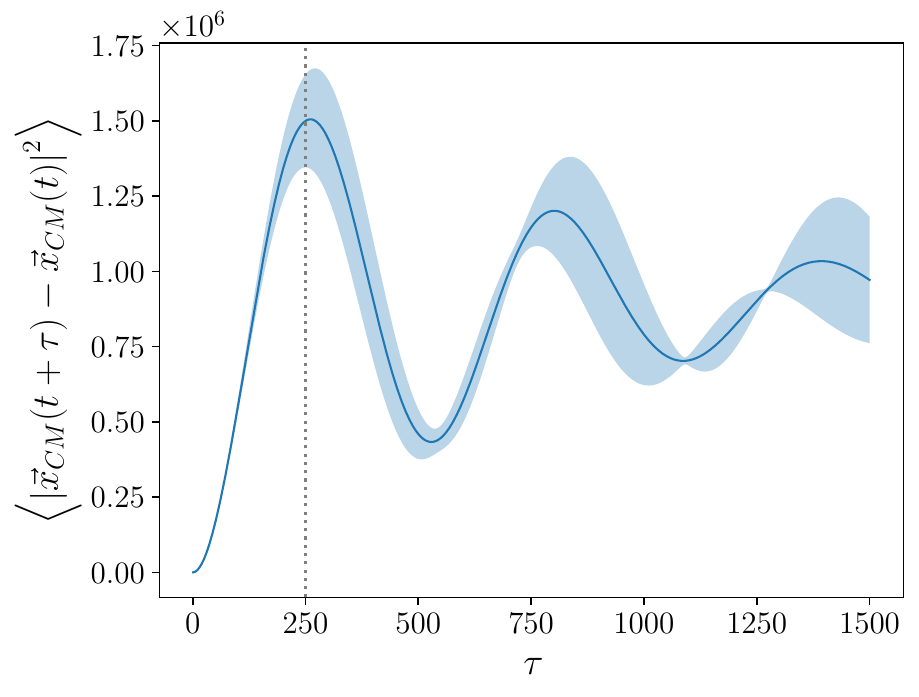}%
  \caption{Mean square displacement for the position of the center of mass
    $\vec{x}_{CM}$ depending on the delayed time $\tau$. The dotted vertical
    line corresponds to the first maximum at
    $\tau=250$. }\label{app:fig:autocorr_xCM}
\end{figure*}

\end{document}